\documentclass[12pt,a4j]{article}
\usepackage[dvips]{graphicx}
\usepackage{enumerate}
\usepackage{amsmath}
\usepackage{amssymb}
\usepackage{amsfonts}

\begin{document}
\baselineskip=7mm
\centerline{\bf A direct method of solution for the Fokas-Lenells derivative }\par
\centerline{\bf nonlinear Schr\"odinger equation: II. Dark soliton solutions}\par
\bigskip
\centerline{Yoshimasa Matsuno\footnote{{\it E-mail address}: matsuno@yamaguchi-u.ac.jp}}\par

\centerline{\it Division of Applied Mathematical Science,}\par
\centerline{\it Graduate School of Science and Engineering} \par
\centerline{\it Yamaguchi University, Ube, Yamaguchi 755-8611, Japan} \par
\bigskip
\bigskip
\leftline{\bf Abstract}\par
\noindent In a previous study (Matsuno Y { \it J. Phys. A: Math. Theor.} {\bf 45} (2012) 23202), we have developed a systematic method for obtaining the
bright soliton solutions of the Fokas-Lenells derivative nonlinear Schr\"odinger equation (FL equation shortly) under  vanishing boundary condition.
In this paper, we apply the method to the FL equation with nonvanishing boundary condition. 
In particular, we deal with a more sophisticated problem on the dark soliton solutions  with a plane wave boundary condition.
We first derive the novel  system of bilinear equations which is reduced from  the FL equation 
through a dependent variable transformation and then construct the general dark $N$-soliton solution of the system,
where $N$ is an arbitrary positive integer.
In the process, a trilinear equation derived from the system of bilinear equations plays an important role. 
As a byproduct, this equation gives the dark $N$-soliton solution of the derivative nonlinear Schr\"odinger equation on the background of a plane wave.
We then investigate the properties of the one-soliton solutions in detail, showing that 
both the dark  and  bright solitons appear on the nonzero  background which reduce to algebraic solitons in  specific limits.
 Last, we perform the asymptotic analysis of the two- and $N$-soliton solutions for large time and clarify their structure and dynamics.

 \par

\newpage
\leftline{\bf  1. Introduction} \par
\bigskip
\noindent The Fokas-Lenells  derivative nonlinear Schr\"odinger (NLS) equation (FL equation shortly) is a completely integrable
nonlinear  partial differential equation (PDE) which has been derived as an
integrable generalization of the NLS equation using  bi-Hamiltonian methods [1].
In the context of nonlinear optics, 
the FL equation models the propagation of nonlinear light  pulses in monomode optical fibers when certain higher-order nonlinear effects are taken into account [2].
We employ the following equation which can be derived from its original version by  a simple change of variables combined with a gauge transformation 
[2]:
$$u_{xt}=u-2{\rm i}|u|^2u_x. \eqno(1.1)$$
Here, $u=u(x,t)$ is a  complex-valued function of $x$ and $t$, 
and subscripts $x$ and $t$ appended to $u$ denote  partial differentiations. 
The complete integrability of the FL equation   has been demonstrated by means of the inverse scattering transform (IST)  method [3].
Especially, a Lax pair and a few conservation laws associated with it have been obtained explicitly using the bi-Hamiltonian structure
and the multisoliton solutions have been derived by applying the dressing method [4].
Another remarkable feature of the FL equation is that it  is the first negative flow of the integrable hierarchy of the derivative NLS equation [2, 5]. \par
In a previous study [6] which is referred to as I hereafter, the two different expressions of the bright $N$-soliton solution of the FL equation have been obtained by a direct method which does not recourse to the IST
and their properties have been explored in detail. Here, we
construct the dark $N$-soliton solution of the FL equation on the background of a plane wave. Explicitly, we consider the boundary condition
$$u \rightarrow \rho\,{\rm exp}\left\{{\rm i}\left(\kappa x-\omega t+\phi^{(\pm)}\right)\right\}, \quad x \rightarrow \pm\infty, \eqno(1.2)$$
where $\rho(>0)$ and $\kappa$ are real constants representing the amplitude and wavenumber, respectively, 
$\phi^{(\pm)}$ are real phase constants
 and the angular frequency $\omega=\omega(\kappa)$ obeys the dispersion relation
$\omega=1/\kappa+2\rho^2.$
Note that the plane wave given in (1.2) is an exact solution of the FL equation. As will be discussed later, the possible values of $\kappa$ must be restricted to assure the existence of the soliton solutions.
A similar problem to that posed in this paper has been studied recently and an explicit formula for the dark $N$-soliton solution
have been presented by  an ingenious application of the B\"acklund transformation between solutions of the FL equation
and the Ablowitz-Ladik hierarchy [7]. Nevertheless, the detailed analysis of the soliton solutions has not  been undertaken as yet. \par
An exact method of solution employed here which is sometimes called the direct method [8] or the bilinear transformation method [9] is a powerful tool for analyzing soliton
equations and differs from the method used in [7]. Once the equation under consideration is transformed to a system of bilinear equations,  the standard  technique 
in the bilinear formalism
is applied to obtain soliton solutions.
A novel feature of the bilinearization of the FL equation  is that one of the bilinear equations can be replaced by a {\it trilinear} equation, as already demonstrated in I. 
The same situation happens in the current dark soliton problem. However, the resulting trilinear equation will be used essentially in the process of performing the proof of the dark $N$-soliton solution. \par
 This paper is organized as follows. In section 2,  we bilinearize the FL equation
 under the boundary condition (1.2). We then show that one of the resulting bilinear equations 
 can be replaced by a trilinear equation.
 In section 3, we present the dark $N$-soliton solution of the bilinear equations.
 It has a simple structure expressed in terms of certain determinants. 
 Subsequently,  we perform the proof of the dark $N$-soliton solution using an elementary theory of determinants 
in which Jacobi's identity  will  play a central role. As already noted, the proof of the trilinear
equation turns out to be a core in the analysis.
In accordance with the relation between the FL  equation and the derivative NLS equation at the level of the Lax representation, we also demonstrate that
the  dark $N$-soliton solution obtained here  yields the dark $N$-soliton solution of the   derivative 
NLS equation by replacing simply 
the time dependence of the solution.  As in the case of the defocusing NLS equation subjected to nonvanishing boundary conditions,
it is necessary for  the  existence of dark solitons that the asymptotic state given by (1.2) must be stable.
Hence, we perform the linear stability analysis of the plane wave solution (1.2)
and provide a criterion for the stability.
In section 4, we first investigate 
the properties of the one-soliton solution in detail. We find that  
depending on the sign of $\kappa$ and that of the real part of the complex amplitude parameter, the solution can be classified into two types, i.e., 
the dark and bright solitons.
The latter soliton may be termed "anti-dark soliton" since the background field is nonzero. However, we use a term "bright soliton" throughout the paper.
We demonstrate that regardless the sign of $\kappa$, the bright soliton has a limiting profile of  algebraic type (or an algebraic bright soliton) 
whereas an algebraic dark soliton appears only if $\kappa<0$.
We then  analyze
the asymptotic behavior of the  two-soliton solution and  derive
 the explicit formulas for the phase shift in terms of the amplitude parameters of solitons.
 In particular,  we address the interaction between a dark soliton and a bright soliton as well as that of two dark solitons. 
Last, the similar asymptotic analysis to that of the two-soliton solution is performed for the general  dark $N$-soliton solution.
Section 5 is devoted to concluding remarks. \par
\bigskip
\leftline{\bf 2. Exact method of solution}\par
\bigskip
\noindent In this section, we develop a direct method of solution for constructing dark soliton solutions of the FL equation (1.1)
under the boundary condition (1.2). 
In particular, we show that it can be transformed to  a system of bilinear equations by introducing the same type of the dependent variable transformation
as that employed in I for the bilinearization of the FL equation under vanishing boundary condition. 
We also demonstrate that this system yields a trilinear equation which will play a crucial role in our analysis.
\par
\bigskip
\leftline{\it 2.1. Bilinearization}\par
\medskip
\noindent The bilinearization of the FL equation (1.1) is established by the following proposition: \par
\medskip
\noindent{\bf Proposition 2.1.}\ {\it By means of the dependent variable transformation
$$u=\rho\,{\rm e}^{{\rm i}(\kappa x-\omega t)}\,{g\over f}, \eqno(2.1)$$
with $\omega=1/\kappa+2\rho^2$, equation (1.1) can be decoupled into the following system of bilinear equations for 
the tau functions $f$ and $g$
$$D_tf\cdot f^*-{\rm i}\rho^2(gg^*-ff^*)=0, \eqno(2.2)$$
$$D_xD_tf\cdot f^*-{\rm i}\rho^2D_xg\cdot g^*+{\rm i}\rho^2D_xf\cdot f^*+2\kappa\rho^2(gg^*-ff^*)=0, \eqno(2.3)$$
$$D_xD_tg\cdot f+{\rm i}\kappa D_tg\cdot f-{\rm i}\omega D_xg\cdot f=0. \eqno(2.4)$$
Here, $f=f(x, t)$ and $g=g(x, t)$ are complex-valued functions of $x$ and $t$,
 and the asterisk appended to $f$ and $g$ denotes complex conjugate and
the bilinear operators $D_x$ and $D_t$ are defined by
$$D_x^mD_t^nf\cdot g=\left({\partial\over\partial x}-{\partial\over\partial x^\prime}\right)^m
\left({\partial\over\partial t}-{\partial\over\partial t^\prime}\right)^n
f(x, t)g(x^\prime,t^\prime)\Big|_{ x^\prime=x,\,t^\prime=t},  \eqno(2.5)$$
where $m$ and $n$ are nonnegative integers.} \par
\bigskip
\noindent {\bf Proof.}  Substituting (2.1) into (1.1) and rewriting the resultant equation in terms of the bilinear operators,
equation (1.1) can be rewritten as
$${1\over f^2}(D_xD_tg\cdot f+{\rm i}\kappa D_tg\cdot f-{\rm i}\omega D_xg\cdot f)$$
$$-{g\over f^3f^*}\bigl\{f^*D_xD_tf\cdot f-2\kappa\rho^2f^2f^*-2{\rm i}\rho^2g^*(g_xf-gf_x+{\rm i}\kappa fg)\bigr\}=0. \eqno(2.6)$$
Inserting the identity 
$$f^*D_xD_tf\cdot f=fD_xD_tf\cdot f^*-2f_xD_tf\cdot f^*+f(D_tf\cdot f^*)_x, \eqno(2.7)$$
which can be verified by direct calculation, into the second term on the left-hand side of (2.6), one modifies it in the form
$${1\over f^2}(D_xD_tg\cdot f+{\rm i}\kappa D_tg\cdot f-{\rm i}\omega D_xg\cdot f)$$
$$-{g\over f^3f^*}\Bigl[f\bigl\{D_xD_tf\cdot f^*-{\rm i}\rho^2D_xg\cdot g^*+{\rm i}\rho^2D_xf\cdot f^*+2\kappa\rho^2(gg^*-ff^*)\bigr\}$$
$$-2f_x\bigl\{D_tf\cdot f^*-{\rm i}\rho^2(gg^*-ff^*)\bigr\}+f\bigl\{D_tf\cdot f^*-{\rm i}\rho^2(gg^*-ff^*)\bigr\}_x\Bigr]=0. \eqno(2.8)$$
By virtue of equations (2.2)-(2.4), the left-hand side of (2.8) vanishes identically. \hspace{\fill}$\Box$ \par
\bigskip
It follows from (2.1) and (2.2) that
$$|u|^2=\rho^2+{\rm i}\,{\partial \over \partial t}\,{\rm ln}\,{f^*\over f}. \eqno(2.9)$$
The above formula gives the modulus of $u$ in terms of the tau function $f$. \par
\bigskip
\leftline{\it  2.2. Trilinear equation} \par
\medskip
\noindent{\bf Proposition 2.2.} {\it The {\it trilinear} equation for $f$ and $g$
$$f^*\left\{g_{xt}f-(f_x-{\rm i}\kappa f)g_t-{\rm i}\left({1\over \kappa}+\rho^2\right)(g_xf-gf_x)\right\}
=f_t^*(g_xf-gf_x+{\rm i}\kappa fg), \eqno(2.10)$$
 is a consequence of the bilinear equations (2.2)-(2.4).} \par
\medskip
\noindent {\bf Proof.} By  direct calculation,  one can show the following trilinear identity among the tau functions $f$ and $g$:
$$f^*\left\{g_{xt}f-(f_x-{\rm i}\kappa f)g_t-{\rm i}\left({1\over \kappa}+\rho^2\right)(g_xf-gf_x)\right\}
-f_t^*(g_xf-gf_x+{\rm i}\kappa fg)$$
$$=f^*(D_xD_tg\cdot f+{\rm i}\kappa D_tg\cdot f-{\rm i}\omega D_xg\cdot f)$$
$$-{g\over 2}\Bigl[\bigl\{D_tf\cdot f^*-{\rm i}\rho^2(gg^*-ff^*)\bigr\}_x+(D_xD_tf\cdot f^*-{\rm i}\rho^2D_xg\cdot g^*+{\rm i}\rho^2D_xf\cdot f^*-2{\rm i}\kappa D_tf\cdot f^*)\Bigr]$$
$$+g_x\bigl\{D_tf\cdot f^*-{\rm i}\rho^2(gg^*-ff^*)\bigr\}. \eqno(2.11)$$
Replacing a term $2{\rm i}\kappa D_tf\cdot f^*$ on the right-hand side of (2.11) by (2.2), the right-hand side becomes zero by (2.2)-(2.4). This yields (2.10).
\hspace{\fill}$\Box$ \par
\bigskip
In view of proposition 2.2, the proof of the dark  $N$-soliton solution is completed if one can prove any three equations among the three bilinear equations (2.2)-(2.4) and a trilinear (2.10).
We will see later in section 3 that the proof of (2.4) is not easy to perform and hence we prove (2.10) instead. \par
\bigskip
\leftline{\bf 3. Dark $N$-soliton solution and its proof}\par
\bigskip
\noindent In this section, we show that the tau functions $f$ and $g$ representing the dark $N$-soliton solution
admit the compact determinantal expressions.
This statement is proved by an elementary calculation using the basic formulas for determinants. We first prove that the proposed dark $N$-soliton 
solution solves the bilinear equations
(2.2) and (2.3) and then the trilinear equation (2.10) in place of (2.4).
 The implication of the equation (2.10) will be discussed  in conjunction with the dark $N$-soliton solution of the derivative NLS equation.
 Last, we perform the linear stability analysis of the plane wave solution (1.2) and provide a criterion for the stability.
\par
\bigskip
\noindent{\it 3.1. Dark $N$-soliton solution} \par
\medskip
\noindent The main result in this paper is given by the following theorem: \par
\noindent{\bf Theorem 3.1.} {\it The dark $N$-soliton solution of the system of bilinear equations (2.2)-(2.4) is expressed by the
 following determinants
$$f=|D|, \eqno(3.1a)$$
$$ g=\begin{vmatrix} D & {\bf z}^T\\ {1\over \rho^2}{\bf z}_t^* & 1\end{vmatrix}=|D|+{1\over \rho^2}\begin{vmatrix} D & {\bf z}^T\\ {\bf z}_t^* & 0\end{vmatrix}. \eqno(3.1b)$$
Here, $D$ is an $N\times N$ matrix and ${\bf z}$ and ${\bf z}_t$  are $N$-component row vectors defined below and 
the symbol $T$  denotes the transpose:
$$D=(d_{jk})_{1\leq j,k\leq N}, \quad d_{jk}=\delta_{jk}+{\kappa-{\rm i}p_j\over p_j+p_k^*}\,z_jz_k^*, 
\quad z_j={\rm exp}\left(p_jx+{\kappa\rho^2\over p_j}t+{1\over p_j+{\rm i}\kappa}\,\tau+\zeta_{j0}\right), \eqno(3.2a)$$
$${\bf z}=(z_1, z_2, ..., z_N), \quad {\bf z}_t=\left({\kappa\rho^2z_1\over p_1}, {\kappa\rho^2z_2\over p_2}, ..., {\kappa\rho^2z_N\over p_N}\right),  \eqno(3.2b)$$ 
where  $p_j$ are complex parameters satisfying the constraints
$$(p_j+{\rm i}\kappa)(p_j^*-{\rm i}\kappa)={1+\kappa\rho^2\over \kappa\rho^2}p_jp_j^*,\quad j=1, 2, ..., N, \eqno(3.2c)$$
 $\zeta_{j0}\ (j=1, 2, ..., N)$ are arbitrary complex parameters, $\delta_{jk}$ is kronecker's delta and $\tau$ is an auxiliary variable.} \par
\bigskip
\par
 The dark $N$-soliton solution is parameterized by $2N$ complex parameters $p_j$ and $\zeta_{j0}\ (j=1, 2, ..., N)$.
 The  parameters $p_j$ determine the amplitude and velocity of the solitons whereas the parameters $\zeta_{j0}$ determine   the 
 phase of the solitons. 
 As opposed to the bright soliton case explored in I, however, the real and imaginary parts of $p_j$ are not independent because of the constraints (3.2c).
 Actually, it may be parameterized either  by the velocity of the $j$th soliton or by  a single angular variable, as will see in section 4.
  An auxiliary variable $\tau$  introduced in (3.2a)  will be used conveniently in performing the proof of (2.10). 
 It can be set to zero after all the calculations have been completed. \par
 \bigskip
 \noindent {\bf Remark 3.1.} The tau function $g$ given by (3.1b) is represented by the determinant of an $(N+1)\times (N+1)$ matrix. It can be rewritten by the
 determinant of  an $N\times N$ matrix. To show this, we multiply the $(N+1)$th  
 column of $g$ by $z_{k,t}^*/\rho^2$ and subtract it from the  $k$th column for $k=1, 2, ..., N$  to obtain
  $$g=\left|\left(\delta_{jk}-{\kappa+{\rm i}p_k^*\over p_j+p_k^*}\,{p_j\over p_k^*}\,z_jz_k^*\right)_{1\leq j,k\leq N}\right|. \eqno(3.3)$$
  Although the tau function from (3.1b) is used in the proof of the dark $N$-soliton solution, an equivalent form (3.3) will be employed in section 4 
  to analyze the structure of the solution. \par
 \bigskip
 \noindent {\bf Remark 3.2.} The complex parameters $p_j$ subjected to the constraints (3.2c) exist only if the condition $\kappa(1+\kappa\rho^2)>0$
 is satisfied, as confirmed easily by putting $p_j=a_j+{\rm i}b_j$ with real $a_j$ and $b_j$. We will show in section 3.6 that this condition is closely related to
 the stability of the plane wave solution of the FL equation. \par
 \bigskip
 \noindent{\it 3.2. Notation and basic formulas for determinants} \par
\medskip
\noindent Before entering into the proof of the dark $N$-soliton solution,
 we first define the  matrices associated with the dark $N$-soliton solution and then provide some basic 
formulas for determinants. Although these formulas  have been used extensively in I, we reproduce them for convenience. \par
The following bordered matrices appear frequently in our analysis:
$$D({\bf a}; {\bf b})=\begin{pmatrix} D & {\bf b}^T\\ {\bf a} & 0\end{pmatrix},\quad
D({\bf a},{\bf b};{\bf c},{\bf d})=\begin{pmatrix} D &{\bf c}^T & {\bf d}^T \\ {\bf a} & 0 &0\\
                                                                            {\bf b} & 0& 0 \end{pmatrix}, \eqno(3.4)$$
where ${\bf a}, {\bf b}, {\bf c}$ and {\bf d} are $N$ component row vectors. 
 Let  $D_{jk}$ be the cofactor of the element $d_{jk}$. The following formulas are well known in the theory of determinants [10]:
$${\partial\over\partial x}|D|=\sum_{j,k=1}^N{\partial d_{jk}\over\partial x}D_{jk}, \eqno(3.5)$$
$$\begin{vmatrix} D & {\bf a}^T\\ {\bf b} & z\end{vmatrix}=|D|z-\sum_{j,k=1}^ND_{jk}a_jb_k,  \eqno(3.6)$$
$$|D({\bf a}, {\bf b}; {\bf c}, {\bf d})||D|= |D({\bf a}; {\bf c})||D({\bf b}; {\bf d})|-|D({\bf a}; {\bf d})||D({\bf b}; {\bf c})|. \eqno(3.7)$$
The formula (3.5) is the differentiation rule of the determinant and (3.6) is the expansion formula for a bordered determinant
with respect to the last row and last column.
The formula (3.7) is Jacobi's identity.   
The proof of lemmas described below is based on the above three formulas as well as a few fundamental
properties of determinants.
\par
\bigskip
\noindent{\it 3.3. Differentiation rules and related formulas} \par
\noindent In terms of the notation (3.4), the tau functions $f$ and $g$ can be written as 
$$f=|D|, \eqno(3.8a)$$
$$\quad g=|D|+{1\over\rho^2}|D({\bf z}_t^*;{\bf z})|. \eqno(3.8b)$$
The differentiation rules of the tau functions with respect to $t$ and $x$ are given by the following formulas: \par
\medskip
\noindent{\bf Lemma 3.1.} \par
$$f_t={\rm i}|D({\bf z}_t^*;{\bf z})| - {1\over \rho^2}|D({\bf z}_t^*;{\bf z}_t)|, \eqno(3.9)$$
$$f_x=-\kappa |D({\bf z}^*;{\bf z})|+{\rm i}|D({\bf z}^*;{\bf z}_x)|, \eqno(3.10)$$
$$f_{xt}={\rm i}\kappa\rho^2|D({\bf z}^*;{\bf z})|-\kappa|D({\bf z}_t^*;{\bf z})|-\kappa|D({\bf z}^*;{\bf z}_t)|
         +{\rm i}|D({\bf z}_t^*;{\bf z}_x)|$$
$$-|D({\bf z}^*,{\bf z}_t^*;{\bf z}_x,{\bf z})|+{\kappa\over \rho^2}|D({\bf z}^*,{\bf z}_t^*;{\bf z},{\bf z}_t)|
-{\rm i\over \rho^2}|D({\bf z}^*,{\bf z}_t^*;{\bf z}_x,{\bf z}_t)|, \eqno(3.11)$$
$$g_t={\rm i}|D({\bf z}_t^*;{\bf z})|+{1\over \rho^2}|D({\bf z}_{tt}^*;{\bf z})|, \eqno(3.12)$$
$$g_x={\rm i}|D({\bf z}_t^*;{\bf z})|+{1\over \rho^2}|D({\bf z}_{t}^*;{\bf z}_x)|
+{\rm i\over \rho^2}|D({\bf z}_t^*,{\bf z}^*;{\bf z},{\bf z}_x)|. \eqno(3.13)$$
\medskip
\noindent {\bf Proof.}  We prove (3.9). Applying formula (3.5) to $f$ given by (3.1) with (3.2a), one obtains
\begin{align}
f_t &=\kappa\rho^2\sum_{j,k=1}^ND_{jk}{\kappa-{\rm i}p_j\over p_jp_k^*}z_jz_k^* \notag \\
    &=-{\rm i}\sum_{j,k=1}^ND_{jk}z_{j}z_{k,t}^* +{1\over \rho^2}\sum_{j,k=1}^ND_{jk}z_{j,t}z_{k,t}^*, \notag
\end{align}
where in passing to the second line, use has been made of the relation $z_{j,t}=(\kappa\rho^2/p_j)z_j$. Referring to  formula (3.6) with $z=0$ and 
taking into account the notation (3.4), the above expression
is equal to the right-hand side of (3.9). Formulas (3.10)-(3.13) can be proved in the same way if one uses (3.5), (3.6) and the relation ${\bf z}_{xt}=\kappa\rho^2{\bf z}$ as well
as some basic properties of determinants. \hspace{\fill}$\Box$ \par
\bigskip
\noindent The complex conjugate expressions of the tau functions $f$ and $g$ and their derivatives are expressed as follows: \par
\medskip
\noindent{\bf Lemma 3.2.}\par
$$f^*=|D|-{\rm i}|D({\bf z}^*;{\bf z})|,  \eqno(3.14) $$
$$f_t^*=-{\rm i}|D({\bf z}^*;{\bf z}_t)|- {1\over \rho^2}|D({\bf z}_t^*;{\bf z}_t)|+{\rm i\over \rho^2}|D({\bf z}_t^*,{\bf z}^*;{\bf z}_t,{\bf z})|, \eqno(3.15)$$
$$g^*=|D|-{\rm i}|D({\bf z}^*;{\bf z})|+{1\over \rho^2}|D({\bf z}^*;{\bf z}_t)|. \eqno(3.16)$$
\medskip
\noindent {\bf Proof.} It follows from (3.2a) that $d_{jk}^*=d_{kj}+{\rm i}z_j^*z_k$ or in  the matrix form,  $D^*= D^T+{\rm i}(z_jz_k^*)_{1\leq j,k\leq N}^T$.
Since $|D^T|=|D|$, one has
$$f^*=|D+{\rm i}(z_jz_k^*)_{1\leq j,k\leq N}|=\begin{vmatrix} D & {\bf z}^T\\ -{\rm i}{\bf z}^* & 1\end{vmatrix}.$$
Applying formula (3.6) to the right-hand side, formula (3.14) follows immediately. Formulas (3.15) and (3.16) can be derived in the same way.
  \hspace{\fill}$\Box$ \par
\bigskip
\noindent{\it 3.4. Proof of the dark $N$-soliton solution}\par
\medskip
\noindent{\it 3.4.1. Proof of (2.2)}\par
\noindent Let 
$$P_1=D_tf\cdot f^*-{\rm i}\rho^2(gg^*-ff^*). \eqno(3.17)$$
Substituting (3.8), (3.9) and (3.14)-(3.16) into (3.17), most terms are canceled, leaving the following three terms
$$P_1={{\rm i}\over \rho^2}\Bigl\{-|D||D({\bf z}_t^*,{\bf z}^*;{\bf z}_t,{\bf z})|+|D({\bf z}^*;{\bf z})||D({\bf z}_t^*;{\bf z}_t)|
-|D({\bf z}_t^*;{\bf z})||D({\bf z}^*;{\bf z}_t)|\Bigr\}.$$
This expression becomes zero by Jacobi's identity. 
 \hspace{\fill}$\Box$ \par
\medskip
\noindent{\it 3.4.2. Proof of (2.3)}\par
\bigskip
\noindent Instead of proving (2.3) directly, we differentiate (2.2) by $x$ and add the resultant expression to (2.3)
and then prove the equation $P_2=0$, where 
$$P_2=f_{xt}f^*-f_xf_t^*-{\rm i}\rho^2(g_xg^*-f_xf^*)+\kappa\rho^2(gg^*-ff^*). \eqno(3.18)$$
Substituting (3.8)-(3.11), (3.13) and (3.14)-(3.16) into (3.18) and rearranging, $P_2$ reduces to
$$P_2={{\kappa}\over \rho^2}\Bigl\{|D||D({\bf z}^*,{\bf z}_t^*;{\bf z},{\bf z}_t)|-|D({\bf z}^*;{\bf z})||D({\bf z}_t^*;{\bf z}_t)|
+|D({\bf z}^*;{\bf z}_t)||D({\bf z}_t^*;{\bf z})|\Bigr\}$$
$$+{{\rm i}\over \rho^2}\Bigl\{-|D||D({\bf z}^*,{\bf z}_t^*;{\bf z}_x,{\bf z}_t)|+|D({\bf z}^*;{\bf z}_x)||D({\bf z}_t^*;{\bf z}_t)|
-|D({\bf z}^*;{\bf z}_t)||D({\bf z}_t^*;{\bf z}_x)|\Bigr\}$$
$$\!+{1\over \rho^2}\Bigl\{\!-|D({\bf z}^*;{\bf z})||D({\bf z}^*,{\bf z}_t^*;{\bf z}_x,{\bf z}_t)|+|D({\bf z}^*;{\bf z}_x)||D({\bf z}_t^*,{\bf z}^*;{\bf z}_t,{\bf z})|
-|D({\bf z}^*;{\bf z}_t)||D({\bf z}^*,{\bf z}_t^*;{\bf z},{\bf z}_x)|\Bigr\}. \eqno(3.19)$$
The first and second terms on the right-hand side of (3.19) vanish by virtue of Jacobi's identity. 
To show that the third term becomes zero as well, we consider the determinantal identity
$$\begin{vmatrix}|D({\bf z}^*;{\bf z})| & |D({\bf z}^*;{\bf z})|& |D({\bf z}_t^*;{\bf z})|\\
                 |D({\bf z}^*;{\bf z}_x)| & |D({\bf z}^*;{\bf z}_x)|& |D({\bf z}_t^*;{\bf z}_x)| \\                        
                 |D({\bf z}^*;{\bf z}_t)| & |D({\bf z}^*;{\bf z}_t)|& |D({\bf z}_t^*;{\bf z}_t)| \end{vmatrix}=0.$$
It is obvious that this determinant is zero since the first two columns coincide. The above assertion
follows immediately by expanding the determinant with respect to the first column and using Jacobi's identity.
Consequently, $P_2=0$.  \hspace{\fill}$\Box$ \par
\bigskip
Before proceeding to the proof of (2.10), we emphasis that the constraints (3.2c) have not been used in the process of the proof of (2.2) and (2.3). 
On the other hand, we find that the proof of (2.4) depends crucially on the constraints. This is an obstacle which has never been encountered 
in performing the proof of the bright $N$-soliton solution (see I).
In conclusion, a direct proof of (2.4) still remains open and hence we shall prove the trilinear equation (2.10) instead. 
It turns out, however that its proof is  found to be unfeasible.
 As we shall now demonstrate,  introduction of an auxiliary variable $\tau$ in the exponential function (3.2) would resolve
this difficulty. \par
\medskip
\noindent{\it 3.4.3. Proof of (2.10)}\par
\bigskip
\noindent We first prepare the  two lemmas to prove (2.10). 
The lemma 3.3 below gives a very simple relation between the partial derivatives $f_t$ and $f_\tau$.
It is to be noted that  the constraints (3.2c) are used only for the proof of this lemma. 
\par
\medskip
\noindent{\bf Lemma 3.3.}\par
$$f_t=(1+\kappa\rho^2)f_\tau, \eqno(3.20a)$$
$$g_t=(1+\kappa\rho^2)g_\tau. \eqno(3.20b)$$
\medskip
\noindent {\bf Proof.} Extracting the factor $z_j$ from the $j$th row and the factor $z_k^*$ from the $k$th column of
the determinant $|D|$, respectively  for $j, k=1, 2, ...,N$,  one can rewrite the tau function $f$ into the form
$$f=\prod_{j=1}^N{\rm e}^{\zeta_j}\left|\left({\rm e}^{-\zeta_j}\delta_{jk}+{\kappa-{\rm i}p_j\over p_j+p_k^*}\right)_{1\leq j,k\leq N}\right|,$$
where
$$\zeta_j=(p_j+p_j^*)x+\kappa\rho^2\left({1\over p_j}+{1\over p_j^*}\right)t+{p_j+p_j^*\over (p_j+{\rm i}\kappa)(p_j^*-{\rm i}\kappa)}\,\tau+\zeta_{j0}+\zeta_{j0}^*,$$
 showing that $f$ can be regarded as a function of $\zeta_j\ (j=1, 2, .., N)$. Thus, differentiation of $f$ with respect to $t$ gives
$$f_t=\sum_{j=1}^N{\partial f\over\partial\zeta_j}{\partial\zeta_j\over\partial t}=\kappa\rho^2\sum_{j=1}^N\left({1\over p_j}+{1\over p_j^*}\right){\partial f\over\partial\zeta_j}. $$
Similarly, one has
$$f_\tau=\sum_{j=1}^N{p_j+p_j^*\over (p_j+{\rm i}\kappa)(p_j^*-{\rm i}\kappa)}{\partial f\over\partial\zeta_j}. $$
The constraints (3.2c) are introduced into the above expression to give
$$f_\tau={\kappa\rho^2\over 1+\kappa\rho^2}\sum_{j=1}^N\left({1\over p_j}+{1\over p_j^*}\right){\partial f\over\partial\zeta_j}={1\over 1+\kappa\rho^2\,}f_t.$$
This completes the proof of (3.20a). Repeating the similar procedure, one can show that the relation (3.20b) holds as well.  \hspace{\fill}$\Box$ \par
\medskip
The lemma 3.4 gives the differentiation rules of $f$ and $g$ with respect to $\tau$: \par
\bigskip
\noindent{\bf Lemma 3.4.}\par
$$f_\tau={\rm i}|D({\bf z}_\tau^*;{\bf z})|, \eqno(3.21)$$
$$f_{\tau}^*=-{\rm i}|D({\bf z}^*;{\bf z}_\tau)|, \eqno(3.22)$$
$$g_\tau={{\rm i}\over \kappa\rho^2}|D({\bf z}_t^*;{\bf z})|+{1\over \rho^2}|D({\bf z}_t^*;{\bf z}_\tau)|, \eqno(3.23)$$
$$g_{x\tau}={\rm i}|D({\bf z}^*;{\bf z})|+{1\over \rho^2}|D({\bf z}_t^*;{\bf z})|+\kappa |D({\bf z}^*;{\bf z}_\tau)|+{{\rm i}\over \kappa\rho^2}|D({\bf z}_t^*;{\bf z}_x)|
            -{{\rm i \kappa}\over \rho^2}|D({\bf z}_t^*;{\bf z}_\tau)|$$
$$-{1\over \kappa\rho^2}|D({\bf z}_t^*,{\bf z}^*;{\bf z},{\bf z}_x)|-{\kappa\over \rho^2}|D({\bf z}_t^*,{\bf z}^*;{\bf z}_\tau,{\bf z})| 
+{{\rm i}\over \rho^2}|D({\bf z}_t^*,{\bf z}^*;{\bf z}_\tau,{\bf z}_x)|. \eqno(3.24)$$
\medskip
\noindent {\bf Proof.}  If one notes the relations
$${\bf z}_{t\tau}=-{{\rm i}\over \kappa}{\bf z}_t+{\rm i}\rho^2{\bf z}_{\tau},\qquad {\bf z}_{x\tau}={\bf z}-{\rm i}\kappa{\bf z}_{\tau},$$
which follows from the definition (3.2a) of $z_j$,
the proof can be done straightforwardly along with the same procedure as that used in the
proof of lemma 3.1 and lemma 3.2. \hspace{\fill}$\Box$ \par
\bigskip
With lemmas (3.2) and (3.3) at hand, we are now ready for starting the proof of (2.10). \par
\medskip
\noindent {\bf Proof of (2.10).} If one replaces the $t$ derivative by the $\tau$ derivative in accordance with (3.20), the trilinear equation (2.10) can   be rewritten in the form
 $$f^*P_3=f_\tau^*P_3^\prime, \eqno(3.25a)$$
 with the bilinear forms $P_3$ and $P_3^\prime$ defined respectively by
 $$P_3=g_{x\tau}f-(f_x-{\rm i}\kappa f)g_\tau-{\rm i\over \kappa}(g_xf-gf_x), \eqno(3.25b)$$
 $$P_3^\prime=g_xf-gf_x+{\rm i}\kappa fg. \eqno(3.25c)$$
 The trilinear equation (3.25) is proved as follows. Substituting (3.8), (3.10), (3.13), (3.23) and (3.24) into (3.25b) and applying Jacobi's identity
to terms multiplied by $|D|$, $P_3$ is simplified considerably. After some elementary calculations, one finds that
$$P_3=\kappa |D({\bf z}^*;{\bf z}_\tau)|\Bigl\{|D|+{1\over\rho^2}|D({\bf z}_t^*;{\bf z})|-{{\rm i}\over \kappa\rho^2}|D({\bf z}_t^*;{\bf z}_x)|\Bigr\}. \eqno(3.26a)$$
Performing the similar calculation for $P_3^\prime$, one obtains
$$P_3^\prime={\rm i}\kappa\Bigl\{|D|-|D({\bf z}^*;{\bf z})|\Bigr\}\Bigl\{|D|+{1\over\rho^2}|D({\bf z}_t^*;{\bf z})|-{{\rm i}\over \kappa\rho^2}|D({\bf z}_t^*;{\bf z}_x)|\Bigr\}. \eqno(3.26b)$$
Taking into account the formulas (3.14) and (3.22), the  expressions (3.26a) and (3.26b) yield (3.25). 
The trilinear equation (3.25) coupled with lemma 3.3 now completes the proof of the trilinear equation (2.10).   \hspace{\fill}$\Box$    \par
\bigskip
\noindent{\it 3.5.  Dark $N$-soliton solution of the derivative NLS equation}\par
\medskip
\noindent In accordance with the fact that the FL  equation is
the first negative flow of the Lax hierarchy of the derivative NLS equation, the spatial part of the Lax pair associated with
the former equation coincides with that of the latter equation with an identification $q=u_x\ [2, 5]$. This observation enables us to obtain
 the dark $N$-soliton solution of the derivative NLS equation
$${\rm i}q_t+q_{xx}+2{\rm i}(|q|^2q)_x=0,\quad q=q(x,t), \eqno(3.27)$$
under the boundary condition
$$q\rightarrow \rho\,{\rm exp}\left\{{\rm i}\left(\kappa x-\omega^\prime t+\psi^{(\pm)}\right)\right\}, \quad x\rightarrow \pm\infty,  \eqno(3.28)$$
where $\omega^\prime=\kappa^2+2\kappa\rho^2$ and $\psi^{(\pm)}$ are real phase constants.  In particular, we establish the following proposition: \par
\medskip
\noindent {\bf Proposition 3.1.} {\it  The dark $N$-soliton solution of the derivative NLS equation (3.27) subjected to the boundary condition (3.28) 
is given in terms of the tau functions $f$ and $h$ by
$$q=\rho\,{\rm e}^{{\rm i}(\kappa x-\omega^\prime t)}\,{hf^*\over f^2}, \eqno(3.29a)$$
with
$$f=|D|, \quad h=|H|. \eqno(3.29b)$$
Here, $D$ and $H$ are $N\times N$ matrices defined respectively by
$$D=(d_{jk})_{1\leq j,k\leq N}, \quad d_{jk}=\delta_{jk}+{\kappa-{\rm i}p_j\over p_j+p_k^*}\,z_jz_k^*, 
\quad z_j={\rm exp}\left[p_jx\!+\!\{{\rm i}p_j^2-2(\kappa+\rho^2)p_j\}t+\zeta_{j0}\right], \eqno(3.30a)$$
$$H=(h_{jk})_{1\leq j,k\leq N}, \quad h_{jk}=\delta_{jk}-{\kappa-{\rm i}p_j\over p_j+p_k^*}\,{p_j\over p_k^*}\,z_jz_k^*, \eqno(3.30b)$$
where  $p_j$ are complex parameters satisfying the constraints
$$p_jp_j^*=\rho^2\{\kappa-{\rm i}(p_j-p_j^*)\},\quad j=1, 2, ..., N, \eqno(3.30c)$$
and  $\zeta_{j0}\ (j=1, 2, ..., N)$ are arbitrary complex parameters.} \par
\bigskip
\noindent {\bf Proof.}   The correspondence between $q$ and $u_x$ mentioned above implies that the relation
$$q=u_x={\partial\over\partial x}\left(\rho\,{\rm e}^{{\rm i}\kappa x}\,{g\over f}\right)=\rho\,{\rm e}^{{\rm i}\kappa x}{1\over f^2}(g_xf-gf_x+{\rm i}\kappa fg),$$
holds at $t=0$.
On the other hand, the expression in the parentheses on the right-hand side is just $P_3^\prime$ defined by (3.25c) and hence it is equal to (3.26b). This fact and (3.14) lead, after
applying the formula (3.6), to
\begin{align}
g_xf-gf_x+{\rm i}\kappa fg &={\rm i}\kappa\Bigl\{|D|-|D({\bf z}^*;{\bf z})|\Bigr\}\Bigl\{|D|+{1\over\rho^2}|D({\bf z}_t^*;{\bf z})|-{{\rm i}\over \kappa\rho^2}|D({\bf z}_t^*;{\bf z}_x)|\Bigr\} \notag \\
                           &={\rm i}\kappa f^*\begin{vmatrix} D & \left(z_j-{{\rm i}p_j\over\kappa}z_j\right)_{1\leq j\leq N}^T \\ \left({\kappa\over p_k^*}z_k^*\right)_{1\leq k\leq N} & 1\end{vmatrix}.     \notag
\end{align}
Multiplying the $(N+1)$th column of the  determinant by $\kappa z_k^*/p_k^*$ and subtracting it from the $k$th column for $k=1, 2, ..., N$, one finds that 
the above expression becomes ${\rm i}\kappa f^*h$. Consequently,
$$q={\rm i}\kappa \rho\,{\rm e}^{{\rm i}\kappa x}\,{f^*h\over f^2}\Bigg|_{t=0}.$$
If one replaces $q$ by ${\rm i}q$ and $\rho$ by $\rho/\kappa$, respectively and introduces the time dependence appropriately, one arrives at (3.29)
  with (3.30). 
The constraints (3.30c) follow from (3.2c) by the above replacement of $\rho$.  
The complex parameters $p_j$ subjected to the constraints (3.30c) exist only if the condition $\kappa+\rho^2>0$ is satisfied.
\hspace{\fill}$\Box$ \par
\medskip
It is instructive to perform the bilinearization of the derivative NLS equation under the boundary condition (3.28).
This provides an alternative way to construct the dark $N$-soliton solution given by proposition 3.1, as we shall see now.
 To this end, following the procedure used in [11, 12], we introduce the gauge transformation
$$q=v\,{\rm exp}\left[{\rm i}\int_{-\infty}^x(\rho^2-|v|^2)dx\right], \eqno(3.31a)$$
as well as the dependent variable transformation for $v$
$$v=\rho\,{\rm e}^{{\rm i}(\kappa x-\omega^\prime t)}\,{h\over f}. \eqno(3.31b)$$
Then, equation (3.27) can be decoupled to the system of bilinear equations for $f$ and $h$
$$D_xf\cdot f^*-{\rm i}\rho^2(hh^*-ff^*)=0, \eqno(3.32)$$
$$D_x^2f\cdot f^*-{\rm i}\rho^2D_xh\cdot h^* +\rho^2(2\kappa+\rho^2)(hh^*-ff^*)=0, \eqno(3.33)$$
$${\rm i} D_th\cdot f+2{\rm i}(\kappa+\rho^2)D_xh\cdot f+D_x^2h\cdot f=0. \eqno(3.34)$$
In view of (3.32), the modulus of $v$ is given in terms of the tau function $f$ by
$$|v|^2=\rho^2+{\rm i}\,{\partial \over \partial x}\,{\rm ln}\,{f^*\over f}, \eqno(3.35)$$
which, combined with (3.31), yields  the formula (3.29).  Note from (3.31a) that $|q|^2=|v|^2$. 
It may be checked by direct computation that the tau functions $f$ and $h$ from (3.29b) with (3.30) satisfy the above bilinear equations. \par
It is important to realize that  we can take the limit $\kappa\rightarrow 0$ for the solution (3.29) since the dispersion relation
is not singular at $\kappa=0$. This gives the  $N$-soliton solution of the derivative NLS equation on a constant background which has been
studied extensively  using various exact methods of solution such as the IST [13-16], B\"acklund transformation [17, 18] and Hirota's direct method [19].
On the other hand, for the  dark $N$-soliton solution given by (2.1), this limiting procedure is not relevant because of
the singular nature of the dispersion relation. \par
Last, we shall briefly describe the properties of the one-soliton solution for the purpose of comparison with those of the one-soliton solution of the FL equation.
Introducing the new real parameters $a_1$ and $b_1$ by $p_1=a_1+{\rm i}b_1$, the square of the modulus of the one-soliton solution from (3.29) and (3.30) with $N=1$ can be written in the form
$$|q_1|^2=\rho^2-{2a_1^2\,{\rm sgn}\, a_1\over \sqrt{a_1^2+(\kappa+b_1)^2}}\,{1\over \cosh\,2(\theta_1+\delta_1)+{(\kappa+b_1){\rm sgn}\, a_1\over \sqrt{a_1^2+(\kappa+b_1)^2}}}. \eqno(3.36a)$$
with
$$\theta_1=a_1(x+c_1t)+\theta_{10},\qquad c_1=2(b_1+\kappa+\rho^2),\qquad {\rm e}^{4\delta_1}={a_1^2+(\kappa+b_1)^2\over 4a_1^2}, \eqno(3.36b)$$
where ${\rm sgn}\, a_1$ denotes the sign of $a_1$, i.e., $a_1=1$ for $a_1>0$ and $a_1=-1$ for $a_1<0$, and $\theta_{10}$ is a real constant.
The constraint (3.30c) then becomes
$$a_1^2+b_1^2=\rho^2(2b_1+\kappa). \eqno(3.37)$$
Using (3.36b) and (3.37), the parameters $a_1$ and $b_1$ are expressed in terms of the velocity $c_1$ of the soliton as
$$a_1^2={1\over 4}\left(c_{\rm max}-c_1\right)\left(c_1-c_{\rm min}\right),\qquad b_1={c_1\over 2}-\kappa-\rho^2, \qquad c_{\rm min}<c_1<c_{\rm max},   \eqno(3.38a)$$
where
$$c_{\rm max}=2(\kappa+2\rho^2)+2\rho\sqrt{\kappa+\rho^2},\qquad  c_{\rm min}=2(\kappa+2\rho^2)-2\rho\sqrt{\kappa+\rho^2}. \eqno(3.38b)$$
One must impose
 the condition $\kappa+\rho^2> 0$ to assure the existence of  soliton solutions. 
 Recall that this condition coincides with a criterion for the stability of the plane wave (3.28) [20].
 We see from (3.36) that
if $a_1>0$, then $|q_1|$ takes the form of a dark soliton whereas if $a_1<0$,  it becomes a bright soliton on a constant background $u=\rho$.
\par
 Let $A_d$ and $A_b$ be the amplitudes of the dark and bright
solitons, respectively with respective to the background. The amplitude-velocity relations follow from (3.36) and (3.38). They read
$$A_d=\rho-\left|\sqrt{c_1-\kappa-2\rho^2}-\sqrt{\kappa+\rho^2}\right|, \eqno(3.39a)$$
$$A_b=\sqrt{c_1-\kappa-2\rho^2}+\sqrt{\kappa+\rho^2}-\rho. \eqno(3.39b)$$
The detailed analysis for the case $\kappa>0$ has been undertaken in [21].  To sum up, the solution has been shown to exhibit the spiky modulation of the amplitude and phase.
It also has been demonstrated that the bright soliton reduces to an algebraic soliton for both limits  $c_1\rightarrow c_{\rm max}$ and   $c_1\rightarrow c_{\rm min}$ whereas the algebraic
dark soliton never exists. In the case $\kappa<0$ which has not been treated in [21],  however, a careful inspection of (3.36) and (3.38) reveals that
the algebraic bright and dark  solitons are produced in the limit  $c_1\rightarrow c_{\rm max}$
and  $c_1\rightarrow c_{\rm min}$, respectively.
The latter new feature is pointed out here for the first time. \par
\bigskip
\noindent {\bf Remark 3.3.}  Using the result obtained in proposition 3.1, we can construct the dark $N$-soliton solution of the modified NLS equation
$${\rm i}q_t+q_{xx}+\mu |q|^2q+{\rm i}\gamma(|q|^2q)_x=0,\quad q=q(x,t), \eqno(3.40)$$
under the boundary condition
$$q\rightarrow \rho\,{\rm exp}\left\{{\rm i}\left(\kappa x-\omega^{\prime\prime} t+\psi^{(\pm)}\right)\right\}, \quad x\rightarrow \pm\infty, \eqno(3.41)$$
where $\omega^{\prime\prime}=\kappa^2-\mu\rho^2+\gamma\kappa\rho^2$ and $\mu$ and $\gamma$ are real constants. To show this, we apply the
gauge transformation
$$q={\rm exp}\left[{\mu\over\gamma}\tilde x+\left({\mu\over\gamma}\right)^2\tilde t\right]\tilde q, \quad x=\tilde x+{2\mu\over\gamma}\,\tilde t, \quad t=\tilde t, \eqno(3.42)$$
to equation (3.40) and see that it can be  recast to the derivative NLS equation ${\rm i}\tilde q_{\tilde t}+\tilde q_{\tilde x\tilde x}+\gamma (|\tilde q|^2\tilde q)_{\tilde x}=0$, which coincides with equation (3.27)
 with the identification $\tilde q=q,\ \tilde x=x,\ \tilde t=t$ and $\gamma=2$.  
The dark $N$-soliton solution of the equation (3.40) then takes the form
$$q= \rho\,{\rm e}^{{\rm i}(\kappa x-\omega^{\prime\prime} t)}\,{{f^\prime}^*h^\prime\over {f^\prime}^2}, \eqno(3.43a)$$
where the tau functions $f^\prime$ and $h^\prime$ are given respectively by
$$f^\prime=\left|\left(\delta_{jk}+{\kappa-{\mu\over\gamma}-{\rm i}p_j\over p_j+p_k^*}\,z_jz_k^*\right)_{1\leq j,k\leq N}\right|,\eqno(3.43b)$$
$$h^\prime=\left|\left(\delta_{jk}-{\kappa-{\mu\over\gamma}-{\rm i}p_j\over p_j+p_k^*}{p_j\over p_k^*}\,z_jz_k^*\right)_{1\leq j,k\leq N}\right|,\eqno(3.43c)$$
with
$$z_j={\rm exp}\left[p_jx\!+\!\{{\rm i}p_j^2-(2\kappa+\gamma\rho^2)p_j\}t+\zeta_{j0}\right]. \eqno(3.43d)$$
The constraints for $p_j$ become
$$p_jp_j^*={\gamma\rho^2\over 2}\left\{\kappa-{\mu\over\gamma}-{\rm i}(p_j-p_j^*)\right\},\quad j=1, 2, ..., N. \eqno(3.44)$$
The complex parameters $p_j$ exist only if the condition $\gamma\left(\kappa-{\mu\over\gamma}+{\gamma\rho^2\over 2}\right)>0$ is satisfied. \par
The following two special cases are worth remarking. The case $\mu=0$ and  $\gamma=2$ reduces to the result given by proposition 3.1. 
On the other hand, in the limit $\gamma\rightarrow 0$ while $\mu$ being fixed, we first
replace $z_j$ by $\sqrt \gamma z_j$ for $j=1, 2, ..., N$ and then take the limit, producing the dark $N$-soliton solution of the
NLS equation. Note, in this limit, that the constraints (3.44) reduce to $p_jp_j^*=-\mu\rho^2/2$ and hence the dark soliton solutions exist only if the condition $\mu<0$ is satisfied. \par
\bigskip
\noindent{\it 3.6.  Stability of the plane wave} \par
\medskip
\noindent  We have  considered the dark solitons on the background of a plane wave $\rho\,{\rm e}^{{\rm i}(\kappa x-\omega t)}$ with $\omega =1/\kappa +2\rho^2$.
It is important to see whether the background field is stable or not against perturbations. If unstable, then dark solitons would not exist, as will be demonstrated in the next section.
To this end, we perform the linear stability analysis of the plane wave. \par
Following the standard procedure, we seek a solution of the form
$$u=(\rho+\Delta\rho)\,{\rm e}^{{\rm i}(\kappa x-\omega t+\Delta\phi)}, \eqno(3.45)$$
where $\Delta\rho=\Delta\rho(x,t)$ and $\Delta\phi=\Delta\phi(x,t)$ are small perturbations. Substituting (3.45) into the FL equation (1.1) and linearizing about the plane wave, we obtain
the system of linear PDEs for $\Delta\rho$ and $\Delta\phi$
$$\Delta\rho_{xt}+\rho(\omega-2\rho^2)\Delta\phi_x-\kappa\rho\Delta\phi_t-4\kappa\rho^2\Delta\rho=0, \eqno(3.46a)$$
$$\rho\Delta\phi_{xt}-(\omega-2\rho^2)\Delta\rho_x+\kappa\Delta\rho_t=0. \eqno(3.46b)$$
Assume the perturbations  of the form ${\rm e}^{{\rm i}(\lambda x-\nu t)}$ with $\lambda$ real and $\nu$ possibly complex 
and substitute them into (3.46) to obtain a homogeneous linear system for  $\Delta\rho$ and $\Delta\phi$
$$(\lambda\nu-4\kappa\rho^2)\Delta\rho+{\rm i}\{\rho\lambda(\omega-2\rho^2)+\kappa\rho\nu\}\Delta\phi=0, \eqno(3.47a)$$
$$-{\rm i}\{(\omega-2\rho^2)\lambda+\kappa\nu\}\Delta\rho+\rho\lambda\nu\Delta\phi=0. \eqno(3.47b)$$
The nontrivial solution exists if $\nu$ satisfies the quadratic equation
$$(\lambda^2-\kappa^2)\nu^2-2(2\kappa\rho^2+1)\lambda\nu-{\lambda^2\over\kappa^2}=0. \eqno(3.48)$$
Solving this equation, we obtain
$$\nu={\lambda\over \lambda^2-\kappa^2}\left[2\kappa\rho^2+1 \pm {1\over\kappa}\sqrt{\lambda^2+4\kappa^3(\kappa\rho^2+1)\rho^2}\right]. \eqno(3.49)$$
Thus,   if the condition
$$\kappa(\kappa\rho^2+1)>0, \eqno(3.50)$$ 
is satisfied, then $\nu$ becomes real for all values of real $\lambda$, implying that the plane wave is neutrally stable.
It is evident that this condition always holds for $\kappa>0$.
For negative $\kappa$, on the other hand, we put $\kappa=-K$ with $K>0$ and see that the stability criterion turns out to be as $K\rho^2>1$. 
Last, we remark that a similar stability analysis has been performed   recently in conjunction with a plane wave solution of the original version of the
FL equation [22, 23].
\par
\bigskip
\noindent{\bf 4. Properties of the  soliton solutions}\par
\bigskip
\noindent In this section, we detail the properties of the  soliton solutions. To this end, 
we first parametrize the complex parameters $p_j$ and $\zeta_{j0}$ by the real quantities $a_j, b_j, \theta_{j0}$ and $\chi_{j0}$ as
$$p_j=a_j+{\rm i}b_j, \qquad \zeta_{j0}=\theta_{j0}+{\rm i}\chi_{j0}, \qquad j=1, 2, ..., N,\eqno(4.1)$$
and introduce the new independent variables $\theta_j$ and $\chi_j$ according to the relations
$$\quad \theta_j=a_j(x+c_jt)+\theta_{j0}, \qquad c_j={\kappa\rho^2\over a_j^2+b_j^2}, \qquad j=1, 2, ..., N. \eqno(4.2a)$$
$$\chi_j=b_j(x-c_jt)+\chi_{j0}, \qquad j=1, 2, ..., N. \eqno(4.2b)$$
In terms of these variables, the variables  $z_j$ defined by (3.2a) are put into the form
$$z_j={\rm e}^{\theta_j+{\rm i}\chi_j}, \qquad j=1, 2, ..., N, \eqno(4.2c)$$
after setting $\tau=0$.
Substituting (4.1) into (3.2c), the constraints for $p_j$ can be rewritten as a quadratic equation for $b_j$
$$b_j^2-2\kappa^2\rho^2b_j+a_j^2-\kappa^3\rho^2=0, \qquad j=1, 2, ..., N.\eqno(4.3)$$
The solution to this equation is found to be as follows:
$$b_j=(\kappa\rho)^2\pm\sqrt{\kappa^3\rho^2(1+\kappa\rho^2)-a_j^2},\qquad j=1, 2, ..., N.\eqno(4.4)$$
We can see from the above expression that the real $b_j\ (j=1, 2, ..., N)$ exist only when the condition
$\kappa^3\rho^2(1+\kappa\rho^2)>0$ is satisfied. 
This coincides with the criterion (3.50) for the stability of the plane wave, as discussed in
section 3.6.  Throughout the analysis, we assume this condition to assure the existence of   soliton solutions. 
It is to be noted from (4.2) and (4.3) that the parameters $a_j$ and $b_j$ are expressed in terms of $c_j$ as
$$a_j^2={\kappa^2\over 4c_j^2}\left(c_{\rm max}-c_j\right)\left(c_j-c_{\rm min}\right), \qquad b_j={1\over 2\kappa c_j}(1-\kappa^2 c_j), \qquad c_{\rm min}<c_j<c_{\rm max}, \eqno(4.5a)$$
where
$$c_{\rm max}={1\over \kappa^2}\left\{1+2\kappa\rho^2+2\sqrt{\kappa\rho^2(1+\kappa\rho^2)}\right\}, \qquad
  c_{\rm min}={1\over \kappa^2}\left\{1+2\kappa\rho^2-2\sqrt{\kappa\rho^2(1+\kappa\rho^2)}\right\}. \eqno(4.5b)$$
The  relations (4.5)  correspond to (3.38) for those of the one-soliton solution of the derivative NLS equation. 
Thus, the dark $N$-soliton solution is characterized by the $N$ velocities $c_j\,(j=1, 2, ..., N)$ and the $2N$ real phase constants $\theta_{j0}$ and $\chi_{j0}\, (j=1, 2, ..., N)$,
the total number of which is $3N$. \par
Another parameterization of the solution is possible if one introduces the angular variables $\gamma_j$ by
$$a_j=\sqrt{\kappa^3\rho^2(1+\kappa\rho^2)}\, \sin\,\gamma_j,  \eqno(4.6a)$$
$$b_j=(\kappa\rho)^2+\sqrt{\kappa^3\rho^2(1+\kappa\rho^2)}\, \cos\,\gamma_j, \quad 0< \gamma_j <2\pi, \quad \gamma_j\not=\pi, \quad j=1, 2, ..., N. \eqno(4.6b)$$
In terms of $\gamma_j$, $p_j$ from (4.1) can be written in the form
$$p_j={\rm i}\left\{(\kappa\rho)^2+\sqrt{\kappa^3\rho^2(1+\kappa\rho^2)}\,{\rm e}^{-{\rm i}\gamma_j}\right\}, \quad j=1, 2, ..., N, \eqno(4.7)$$
and  the velocity $c_j$  of the $j$th soliton given in (4.2a) is  expressed as
$$c_j={1\over \kappa^2\{1+4\kappa\rho^2(1+\kappa\rho^2)\,\sin^2\gamma_j\}}\left\{1+2\kappa\rho^2-2\,{\rm sgn}\,\kappa\,\sqrt{\kappa\rho^2(1+\kappa\rho^2)}\,\cos\,\gamma_j\right\}. \eqno(4.8)$$
It follows from the above parametric representation that $p_j$ lies on the circle of radius $\sqrt{\kappa^3\rho^2(1+\kappa\rho^2)}$ centered at ${\rm i}(\kappa\rho)^2$
in the complex plane.
Let us  first describe  the properties of the  one- and two-soliton solutions and then address the general  $N$-soliton solution. \par
\medskip
\noindent{\it 4.1. One-soliton solution}\par
\medskip
\noindent The tau functions $f=f_1$ and $g=g_1$ for the one-soliton solution follows from (3.1)-(3.3) with $N=1$. They read
$$f_1=1+{\kappa-{\rm i}p_1\over p_1+p_1^*}\,z_1z_1^*,\qquad g_1=1-{\kappa+{\rm i}p_1^*\over p_1+p_1^*}{p_1\over p_1^*}\,z_1z_1^*. \eqno(4.9)$$
The one-soliton solution $u_1$  follows from (2.1) with (4.9), yielding
$$u_1=\rho\,{\rm e}^{{\rm i}(\kappa x-\omega t)}\,{1-{\kappa+b_1+{\rm i}a_1\over 2a_1}\,{a_1+{\rm i}b_1\over a_1-{\rm i}b_1}\,{\rm e}^{2\theta_1}\over 1+{\kappa+b_1-{\rm i}a_1\over 2a_1}\,{\rm e}^{2\theta_1}}.
 \eqno(4.10)$$
The above expression can be put into the form 
$$u_1=|u_1|\,{\rm e}^{{\rm i}(\kappa x-\omega t)}{\rm exp}\left\{{\rm i}\left(\phi+\phi^{(+)}\right)\right\}, \eqno(4.11)$$
where   the square of the modulus of $u_1$ is represented by
$$|u_1|^2=\rho^2-{2a_1^2c\,{\rm sgn}(\kappa a_1)\over \sqrt{a_1^2+(\kappa+b_1)^2}}\,{1\over \cosh\,2(\theta_1+\delta_1)+{(\kappa+b_1)\,{\rm sgn}\,a_1\over \sqrt{a_1^2+(\kappa+b_1)^2}}}, \qquad c=|c_1|, \eqno(4.12a)$$
with
$$\theta_1=a_1(x+c_1t)+\theta_{10},\qquad c_1={\kappa\rho^2\over a_1^2+b_1^2},\qquad {\rm e}^{4\delta_1}={a_1^2+(\kappa+b_1)^2\over 4a_1^2}, \eqno(4.12b)$$
and the tangent of the phase $\phi$ and $\phi^{(+)}$ being given respectively by 
$$\tan\, \phi={\{a_1^2+b_1(\kappa+b_1)\}\,\cosh\,2(\theta_1+\delta_1)+b_1\,{\rm sgn}\, a_1\sqrt{a_1^2+(\kappa+b_1)^2}\over \kappa a_1\,\sinh\,2(\theta_1+\delta_1)},  \eqno(4.13a)$$
$$\tan\,\phi^{(+)}={a_1^2+b_1(\kappa+b_1)\over \kappa a_1}. \eqno(4.13b)$$
It can be confirmed by direct substitution that (4.11) indeed  satisfies the FL equation.
The one-soliton solution (4.10) is a one-parameter family of solutions. 
The parameterization in terms of $a_1$ will be employed in classifying the soliton solutions. The parameters $c_1$ and $b_1$ are then expressed by $a_1$.  See (4.2a) and (4.4)
whereas the parameters $\rho$ and $\kappa$ are fixed by the boundary condition (1.2). 
 The relation (4.5) will be used conveniently when considering the generation of algebraic solitons in the limit $|a_1|\rightarrow 0$.
The form of $|u_1|$ from (4.12) reveals that
If $\kappa a_1>0$, then $|u_1|$ takes the form of a dark soliton whereas if $\kappa a_1<0$,  it becomes a bright soliton on a constant background $u=\rho$.
Note from (4.12) that the width of the soliton may be defined by $(2|a_1|)^{-1}$.
The net change of the phase caused by the effect of nonlinear modulation is given by (4.13).  Roughly speaking,
the phase $\phi$ behaves like a step function as a function of $\theta_1$. Specifically, a rapid change of the phase occurs 
in the vicinity of the center position of the soliton ($\theta_1=-\delta_1$), yielding a phase difference $\pi$ (or $-\pi$).
As a result, the phase of $u_1$ changes by a quantity $2\phi^{(+)}$ as $\theta_1$ varies from $-\infty$ to $+\infty$, where $\phi^{(+)}$ is given by (4.13b). \par
Let us  classify the one-soliton solutions in accordance with the sign of $\kappa$. 
We consider the two cases, i.e., case 1 ($\kappa>0, a_1\lessgtr 0$) and case 2 ($\kappa<0, a_1\lessgtr 0$) separately. 
For each sign of $\kappa$, both dark and bright solitons  arise, as we shall show now. \par

\medskip
\noindent {\it 4.1.1.  Case 1: $\kappa>0$ }\par
\medskip
\noindent  In this case, the velocity $c_1$ of the soliton is positive, as evidenced from (4.12b).
Let $A_d$ and $A_b$ be the amplitudes of the dark and bright
solitons, respectively with respective to the background. We then find from (4.5) and (4.12) that
\begin{align}
A_d &= \rho-\sqrt{\rho^2-2c_1\left\{\sqrt{a_1^2+(\kappa+b_1)^2}-(\kappa+b_1)\right\}} \notag \\
     &= \rho-{1\over\sqrt{\kappa}}\left|\kappa\sqrt{c}-\sqrt{1+\kappa\rho^2}\right|, \qquad  a_1>0, \quad c_1=c>0,\tag{4.14}
\end{align}
\begin{align}
A_b &=\sqrt{\rho^2+2c_1\left\{\sqrt{a_1^2+(\kappa+b_1)^2}+(\kappa+b_1)\right\}}-\rho \notag \\
    &={1\over\sqrt{\kappa}}\left(\kappa\sqrt{c}+\sqrt{1+\kappa\rho^2}\right)-\rho, \qquad  a_1<0, \quad c_1=c>0,\tag{4.15}
\end{align}
where $c\equiv |c_1|$ lies in the interval $c_{\rm min}<c<c_{\rm max}$  with $c_{\rm max}$ and $c_{\rm min}$ being given by (4.5b).
Note from (4.5a) that $\kappa+b_1=(1+\kappa^2c_1)/(2\kappa c_1)>0$ for $\kappa>0$ and $c_1>0$. 
This estimate will be used to judge the existence of algebraic solitons in the limit of  infinite width.
\par
\begin{figure}[t]
\begin{center}
\includegraphics[width=10cm]{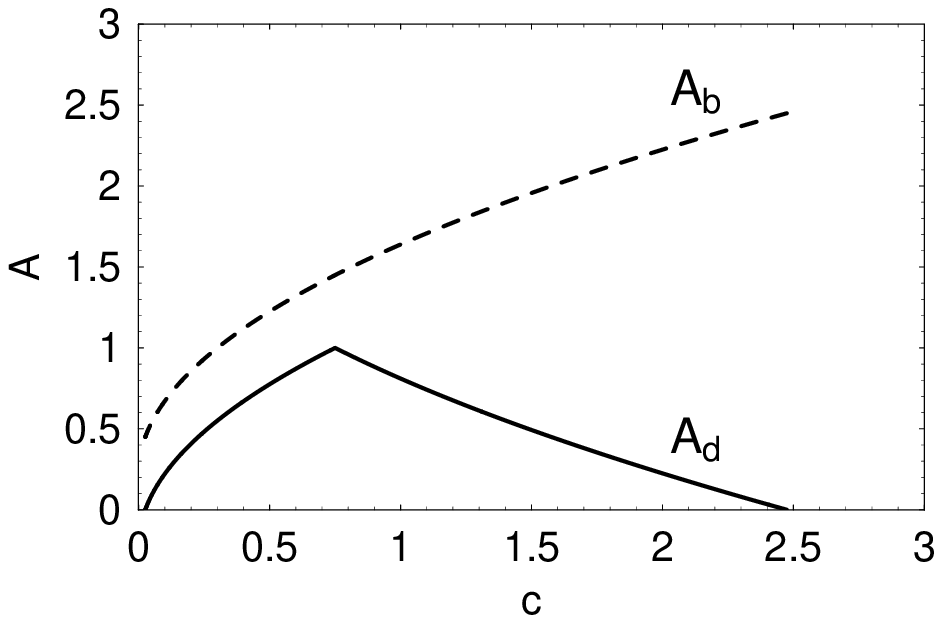}
\end{center}
{\bf Figure 1.} Amplitude-velocity relation for the dark soliton $A_d$ (solid line) and bright soliton $A_b$ (broken line) for $\rho=1$ and $\kappa=2$. \par
\end{figure}

Figure 1 plots the dependence of the amplitudes $A=A_d$ and $A=A_b$ on the velocity $c=|c_1|$ for $\rho=1$ and $\kappa=2$. \par
\medskip
\begin{figure}[t]
\begin{center}
\includegraphics[width=10cm]{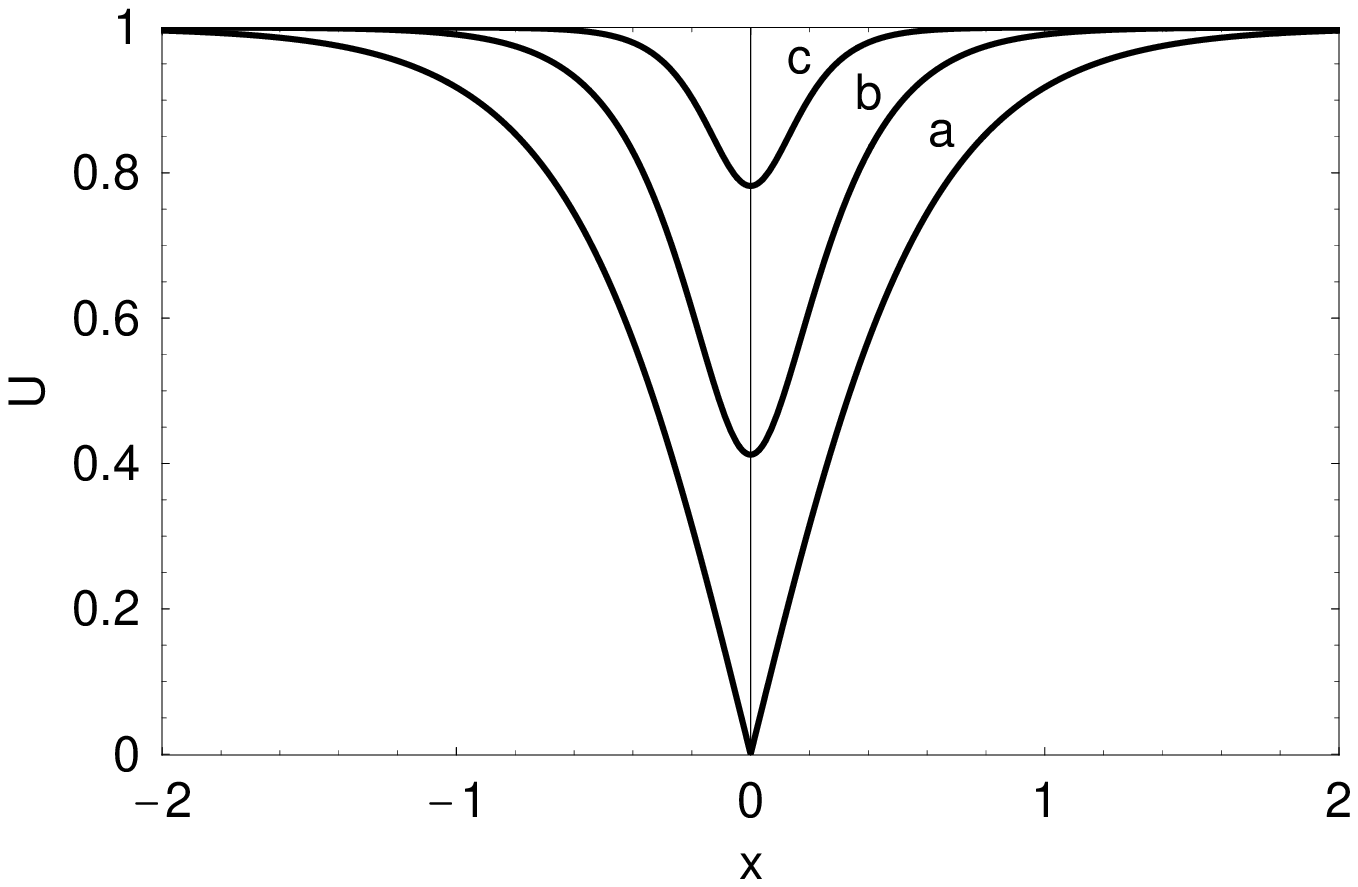}
\end{center}
{\bf Figure 2.} Profile of the amplitude of the dark soliton $U=|u_1|$ at $t=0$. a: $c=c_0=0.75$, b: $c=0.33$, c: $c=0.098$. The profile a is a black soliton.
\end{figure}
\noindent {\it (i) Dark soliton: $a_1>0$}\par
\medskip
\noindent  As seen from figure 1, the amplitude $A_d$ of the dark soliton becomes an increasing function of the velocity $c$
in the interval $c_{\rm min}< c\leq c_0$ and a decreasing function in the interval $c_0<c< c_{\rm max}$, where
$c_{\rm max}\, (\gamma_1=\pi)$ and $c_{\rm min}\,(\gamma_1=0)$ are given by (4.5b) and  a critical velocity $c_0$ and the corresponding angle $\gamma_0$ by
$$c_0={1+\kappa\rho^2\over \kappa^2}, \qquad {\rm at}\quad \gamma_1=\gamma_0=\cos^{-1}\left[-{(\kappa\rho^2)^{1\over 2}(3+2\kappa\rho^2)\over 2(1+\kappa\rho^2)^{3\over 2}}\right], \qquad (0<\gamma_0<\pi). \eqno(4.16)$$
In the present numerical example ($\rho=1, \kappa=2$), $c_{\rm min}=0.025, c_0=0.75, c_{\rm max}=2.47$.  
The above observation shows that  in the interval $c_0<c<c_{\rm max}$, a small dark soliton propagates faster than a large dark soliton. 
A similar behavior has also been found in I for the bright soliton solutions of the FL equation with zero background. \par
Figure 2 depicts
the profile of $U=|u_1|$ at $t=0$ for three different values of $c$, i.e., a: $c=c_0=0.75 (\gamma_1=\gamma_0=0.90\pi)$, b: $c=0.33 (\gamma_1=5\pi/6)$, c: $c=0.098 (\gamma_1=2\pi/3)$
with the parameters $\rho=1, \kappa=2, \theta_{10}=-\delta_1$ and $\chi_{10}=0$. When $c=c_0$, the amplitude of the dark soliton attains the maximum value $A_d=\rho$. 
See figure 2 a. It then turns out
that the intensity of the soliton center falls to zero. Such a soliton is well-known in the field of nonlinear optics. It is
sometimes called a {\it black} soliton.
For this specific value of $c$, one finds from  (4.5), (4.12)  and (4.13) that
$$a_1={\kappa^{3\over 2}\rho(4+3\kappa\rho^2)^{1\over 2}\over 2(1+\kappa\rho^2)}, \quad b_1=-{(\kappa\rho)^2\over  2(1+\kappa\rho^2)}, \eqno(4.17a)$$
$$ {\rm e}^{4\delta_1}={\kappa^2\over 4a_1^2}, \qquad \tan\,\phi=-{b_1\over a_1}\, \tanh(\theta_1+\delta_1),\qquad \tan\,\phi^{(+)}=-{b_1\over a_1}. \eqno(4.17b)$$
The profile of $|u_1|^2$ from (4.12) then becomes
$$|u_1|^2=\rho^2\left[1-{4+3\kappa\rho^2\over 2(1+\kappa\rho^2)}{1\over \cosh\,2(\theta_1+\delta_1)+{2+\kappa\rho^2\over 2(1+\kappa\rho^2)}}\right]. \eqno(4.18)$$
As confirmed easily from the above expression, the minimum value of $|u_1|$ is zero at $\theta_1=-\delta_1$. 
 The algebraic dark soliton may be produced from (4.12) by taking  the limit  $a_1\rightarrow +0$. However, as already noticed, the value of $\kappa+b_1$ is positive 
 so that $|u_1|$ tends simply to a constant value $\rho$.
 Hence, this limiting procedure is irrelevant for the  dark soliton solution under consideration, indicating that the  algebraic dark soliton does not exist for $\kappa>0$ and $a_1>0$.
\par
\begin{figure}[t]
\begin{center}
\includegraphics[width=10cm]{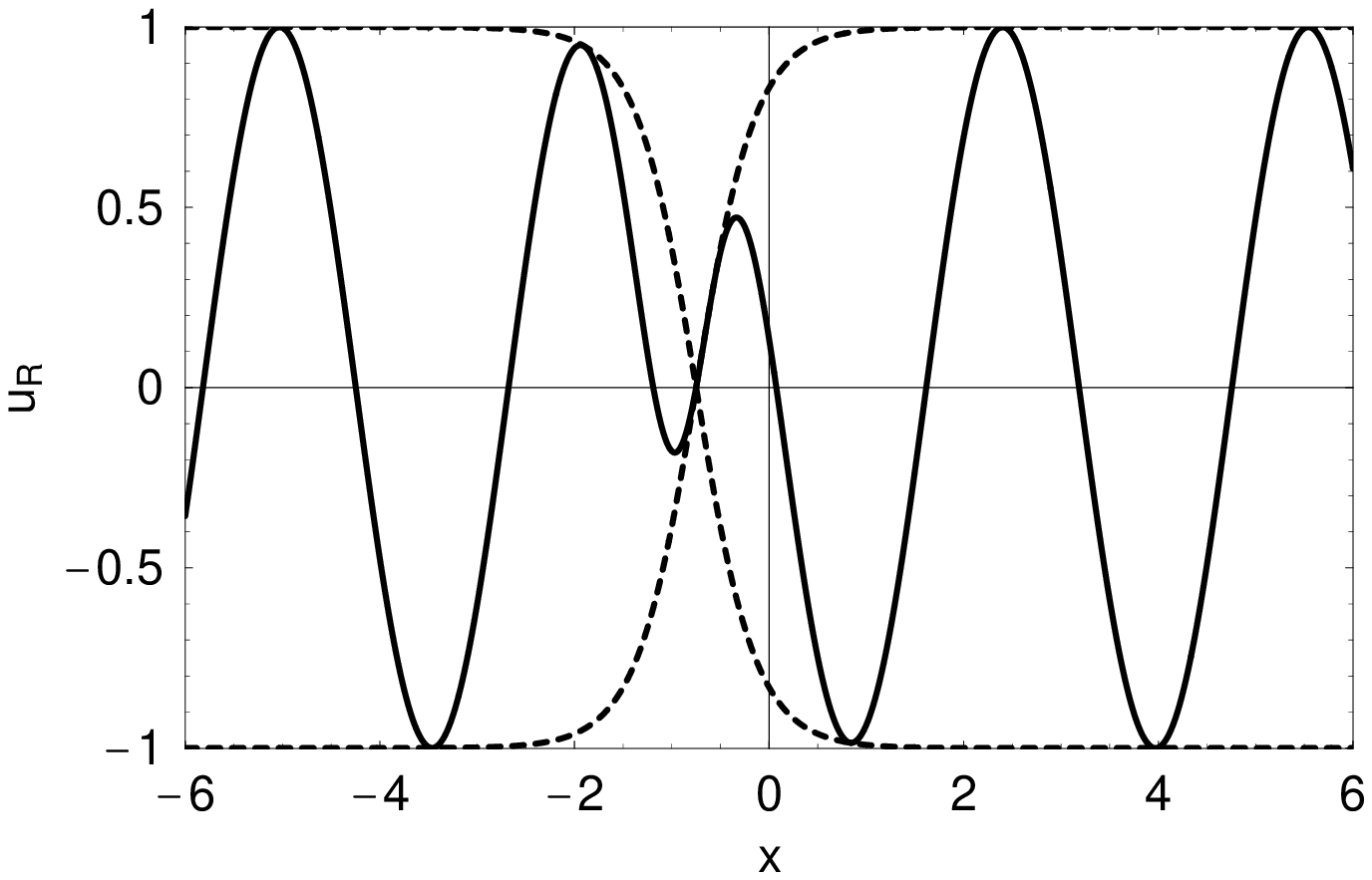}
\end{center}
\centerline{{\bf Figure 3.} Profile of a black soliton $u_{\rm R}={\rm Re}\,u_1$ at $t=1$.} \par
\end{figure}
 Figure 3 shows the profile of $u_{\rm R}={\rm Re}[u_1]$ at $t=1$ for the black soliton. The broken line indicates $\pm |u_1|$ (see figure 2 a). 
One can see that the dark soliton exhibits   phase modulations near the center position of the soliton. This peculiar feature is in striking contrast
to the bright soliton solution of the NLS equation for which no phase modulation occurs. A similar behavior has been observed for both dark and bright soliton solutions of the
derivative NLS equation with the background  of a plane wave [21, 24]. 
\medskip
\par
\noindent{\it (ii)  Bright soliton: $a_1<0$}\par
\noindent Figure 4 depicts the profile of the bright soliton $U=|u_1|$ at $t=0$ for three different values of $c$, i.e., a: $c=2.47 (\gamma_1=1.001\pi)$, b: $c=0.73 (\gamma_1=1.1\pi)$, c: $c= 0.025 (\gamma_1=1.999\pi)$
with $\rho=1$ and $\kappa=2$.
    The feature of  the bright soliton differs substantially from that of the dark soliton. To be specific, the amplitude of the bright soliton always becomes an increasing function of the velocity (see figure 1).
It takes the maximum value at $c=c_{\rm max} (\gamma_1\rightarrow\pi+0, a_1\rightarrow -0)$ and the minimum value at $c=c_{\rm min} (\gamma_1\rightarrow 2\pi-0, a_1\rightarrow -0)$. At these limiting values of the velocity,
the algebraic soliton is  produced from the soliton  of hyperbolic type. Indeed, if we put $\theta_{10}=a_1x_0-\delta_1$ in (4.10) and (4.12) with $x_0$ being a real constant 
and then  take the limit $a_1\rightarrow -0$, we find
$$u_1=\rho\,{\rm e}^{{\rm i}(\kappa x-\omega t)}\,{x+ct+x_0-{\rm i}\,{2\kappa+b_1\over 2b_1(\kappa+b_1)} \over x+ct+x_0-{\rm i}\,{1\over 2(\kappa+b_1)}}, \eqno(4.19a)$$
$$|u_1|^2=\rho^2+{2\kappa c^2\over 1+\kappa^2c}\,{1\over (x+ct+x_0)^2+\left({\kappa c\over 1+\kappa^2c}\right)^2},\eqno(4.19b)$$
where $b_1=(1-\kappa^2c)/2\kappa c$ by (4.5a) and $c=c_{\rm max}$ or $c_{\rm min}$. 
Note from (4.12b) that $b_1^2=\kappa\rho^2/c$ when $a_1\rightarrow -0$. One can see that the algebraic soliton has no free parameters except a phase constant $x_0$
since the velocity $c$ is determined by $\rho$ and $\kappa$ which are fixed by the boundary condition. \par
To derive (4.19a) from (4.10), we use the following expansion formulas for small $a_1$:
$${\rm e}^{2\theta_1} ={2|a_1|\over \sqrt{a_1^2+(\kappa+b_1)^2}}\,{\rm e}^{2a_1(x+ct+x_0)} \sim {2|a_1|\over |\kappa+b_1|}\Big\{1+2a_1(x+ct+x_0)+O(a_1^2)\Big\}, \eqno(4.20a)$$
$${\kappa+b_1-{\rm i}a_1\over 2a_1}\,{\rm e}^{2\theta_1} \sim {\rm sgn}\, a_1\, {\rm sgn}(\kappa+b_1)\left[1+a_1\left\{2(x+ct+x_0)-{\rm i}\,{1\over \kappa+b_1}\right\}+O(a_1^2)\right], \eqno(4.20b)$$
$${\kappa+b_1+{\rm i}a_1\over 2a_1}{a_1+{\rm i}b_1\over a_1-{\rm i}b_1}\,{\rm e}^{2\theta_1}$$
$$\sim -{\rm sgn}\, a_1\, {\rm sgn}(\kappa+b_1)\left[1+a_1\left\{2(x+ct+x_0)-{\rm i}\,{2\kappa+b_1\over b_1(\kappa+b_1)}\right\}+O(a_1^2)\right]. \eqno(4.20c)$$
\begin{figure}[t]
\begin{center}
\includegraphics[width=10cm]{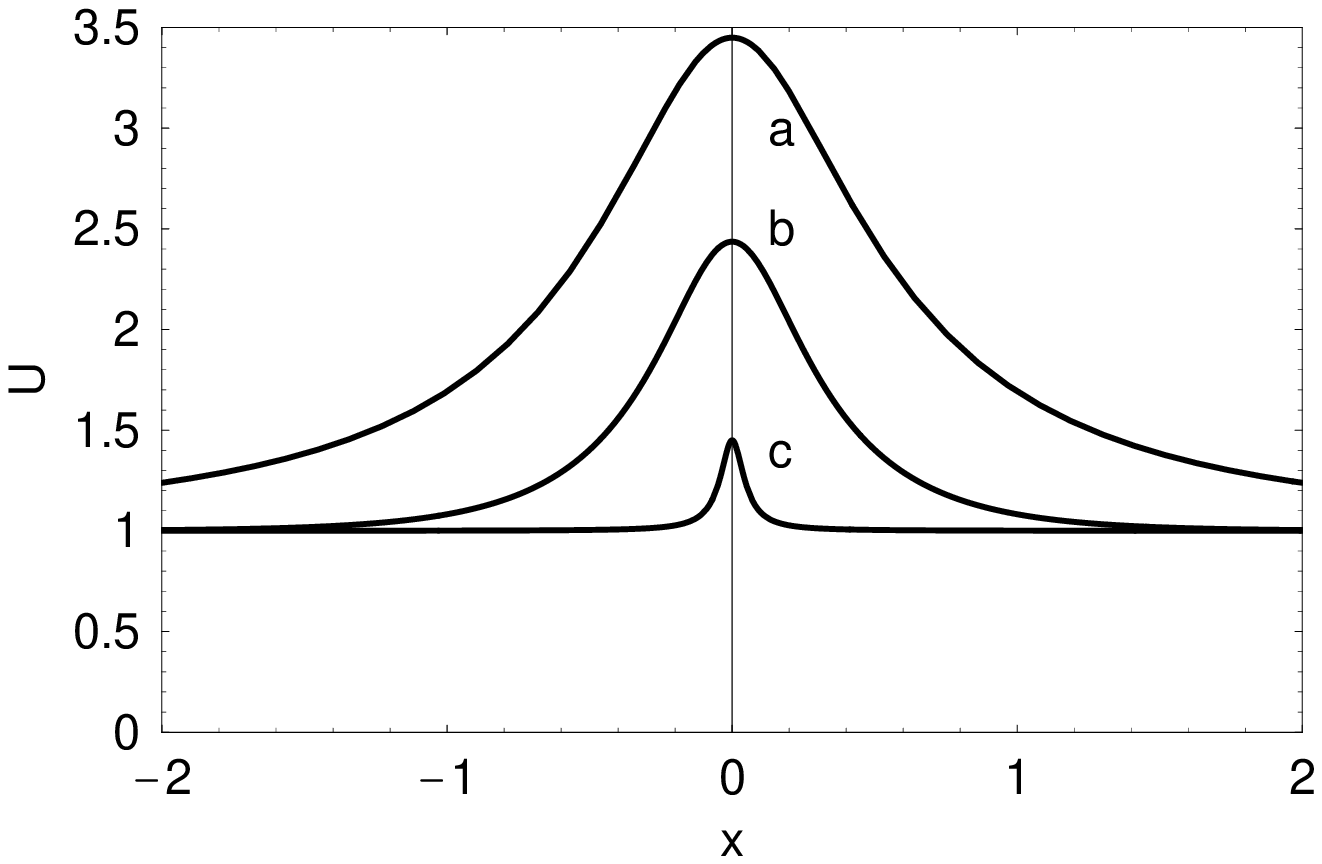}
\end{center}
{\bf Figure 4.} Profile of the amplitude of the bright soliton $U=|u_1|$ at $t=0$. a: $c=2.47$, b: $c=0.73$, c: $c=0.025$. The profiles  a and c are algebraic solitons.\par
\end{figure}
\noindent Because of the inequalities $a_1<0$ and $\kappa+b_1>0$ in the current problem, one finds that 
the condition ${\rm sgn}\, a_1\, {\rm sgn}(\kappa+b_1)=-1$ is satisfied, which yields (4.19a) by taking the limit  $a_1\rightarrow -0$ for (4.10).
Actually, under the above condition, the leading-order terms of the denominator and numerator of (4.10) turn out to be of order $a_1$. Consequently, the expression (4.10) has a
limiting form (4.19a) in the zero limit of $a_1$. On the other hand, the expression (4.19b) follows  either directly from (4.19a) or from (4.12) by performing the similar limiting procedure. 

\begin{figure}[t]
\begin{center}
\includegraphics[width=10cm]{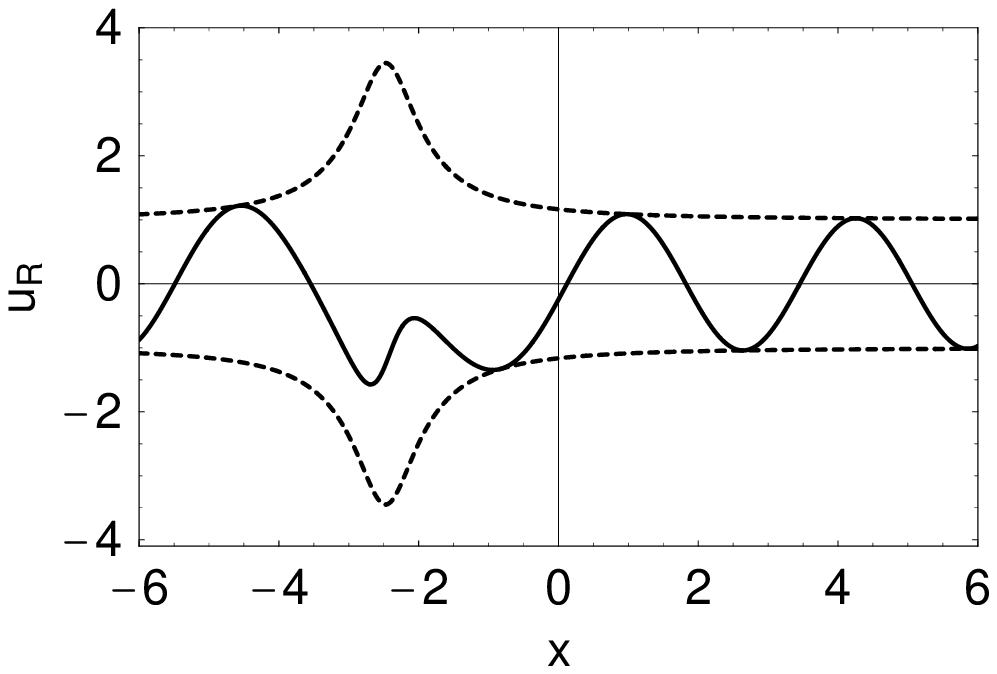}
\end{center}
\centerline{{\bf Figure 5.} Profile of an algebraic  bright soliton $u_{\rm R}={\rm Re}\,u_1$ at $t=1$.}\par
\end{figure}
 A representative profile of the  algebraic bright soliton $U=|u_1|$ at $t=0$ and the corresponding
 profile of $u_{\rm R}={\rm Re}\,u_1$ at $t=1$ are  shown in figure 4 a and figure 5, respectively.  \par
 The novel feature of the bright soliton mentioned above deserves a few comments. First, the amplitude of the bright soliton tends to a finite value when its width tends to infinity, as opposed to the
 behavior of the dark soliton discussed just before for which 
 the amplitude  becomes zero in this limit. 
 Second, the FL equation has an infinite number of conservation laws [3]. Among them, we evaluate the conserved quantity $I=\int_{-\infty}^\infty(|u_x|^2-\kappa^2\rho^2)dx$ for the one-soliton
 solution (4.10). This quantity may be termed the energy of the soliton in accordance with the correspondence between the solution $u$ of the FL equation and the solution $q$ of the derivative NLS
 equation. 
 Using the relation $(|u_x|^2)_t=(|u|^2)_x$ which follows directly from the FL equation, we obtain 
  $$I=-4\,{\rm sgn}\, a_1\,\tan^{-1}\left[{1\over |a_1|}\left\{\sqrt{a_1^2+(\kappa+b_1)^2}-(\kappa+b_1){\rm sgn}\, a_1\right\}\right]. $$
   We find from this expression that in the limit of infinite width $|a_1|\rightarrow 0$, 
  $I$ becomes zero for the dark soliton $(a_1>0)$ and
    tends to a finite value $2\pi$ for the bright soliton $(a_1<0)$.
  See also an analogous calculation for the  bright soliton solution of the derivative NLS equation with zero background [25].
  \par
  \medskip
  \noindent {\it 4.1.2. Case 2: $\kappa<0$} \par
  \noindent For negative $\kappa$, the expressions of the amplitude for the dark and bright solitons are given respectively by
  \begin{align}
A_d &= \rho-\sqrt{\rho^2+2c_1\left\{\sqrt{a_1^2+(\kappa+b_1)^2}+(\kappa+b_1)\right\}} \notag \\
     &= \rho-{1\over\sqrt{K}}\left|K\sqrt{c}-\sqrt{K\rho^2-1}\right|, \qquad  a_1<0, \quad c_1=-c<0,\tag{4.21}
\end{align}
\begin{align}
A_b &=\sqrt{\rho^2-2c_1\left\{\sqrt{a_1^2+(\kappa+b_1)^2}-(\kappa+b_1)\right\}}-\rho \notag \\
    &={1\over\sqrt{K}}\left(K\sqrt{c}+\sqrt{K\rho^2-1}\right)-\rho, \qquad  a_1>0, \quad c_1=-c<0,\tag{4.22}
\end{align}
 where $K=-\kappa$ is a positive wavenumber and the velocity $c$ lies in the interval $c_{\rm min}^\prime<c<c_{\rm max}^\prime$ with
 $$c_{\rm max}^\prime={1\over K^2}\left\{2K\rho^2\!-\!1\!+2\sqrt{K\rho^2(K\rho^2\!-\!1)}\right\},\
  c_{\rm min}^\prime={1\over K^2}\left\{2K\rho^2\!-\!1\!-\!2\sqrt{K\rho^2(K\rho^2\!-\!1)}\right\}. \eqno(4.23)$$
  Recall that the condition $K\rho^2-1>0$ must be imposed to assure the existence of the soliton solutions. \par
  \begin{figure}[t]
\begin{center}
\includegraphics[width=10cm]{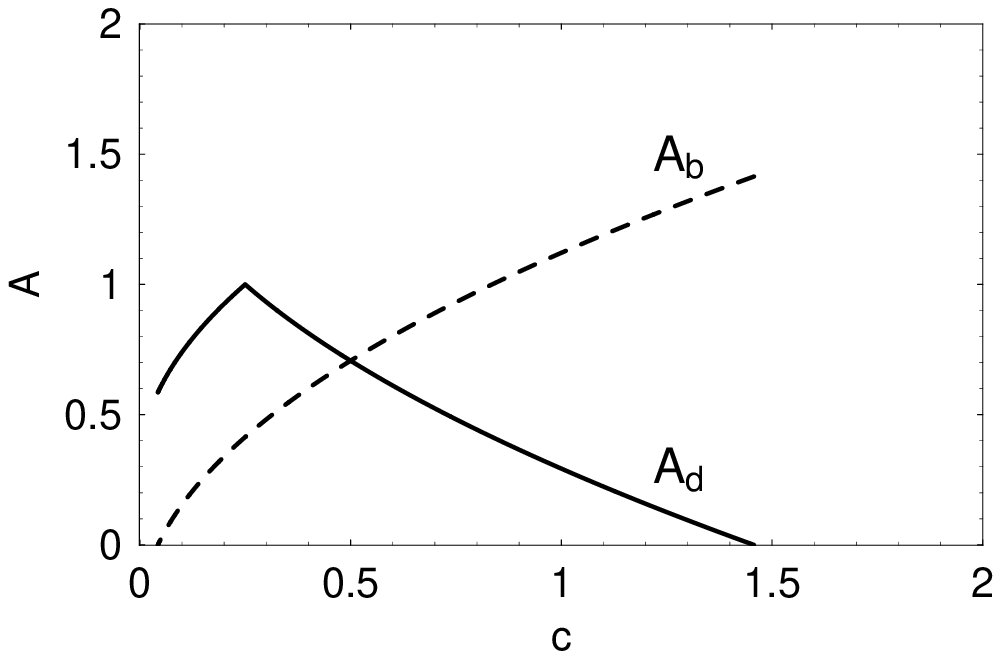}
\end{center}
{\bf Figure 6.} Amplitude-velocity relation for the dark soliton $A_d$ (solid line) and bright soliton $A_b$ (broken line) for $\rho=1$ and $\kappa=-2$. \par
\end{figure}
 Figure 6 plots the dependence of the amplitudes $A=A_d$ and $A=A_b$ on the velocity $c=|c_1|$ for $\rho=1$ and $\kappa= -2$. 
 When compared with figure 1 for $\kappa>0$, there appear several different features for $\kappa<0$.
 In particular, the algebraic {\it dark} soliton would arise in the limit $c\rightarrow c_{\rm min}^\prime$ since in this limit, the amplitude $A_d$
 tends to a finite value. In addition, the algebraic bright soliton exists only in the limit $c\rightarrow c_{\rm max}^\prime$.  We now proceed to the detailed description of
 the soliton solutions.  \par
   \medskip
      \begin{figure}[t]
\begin{center}
\includegraphics[width=10cm]{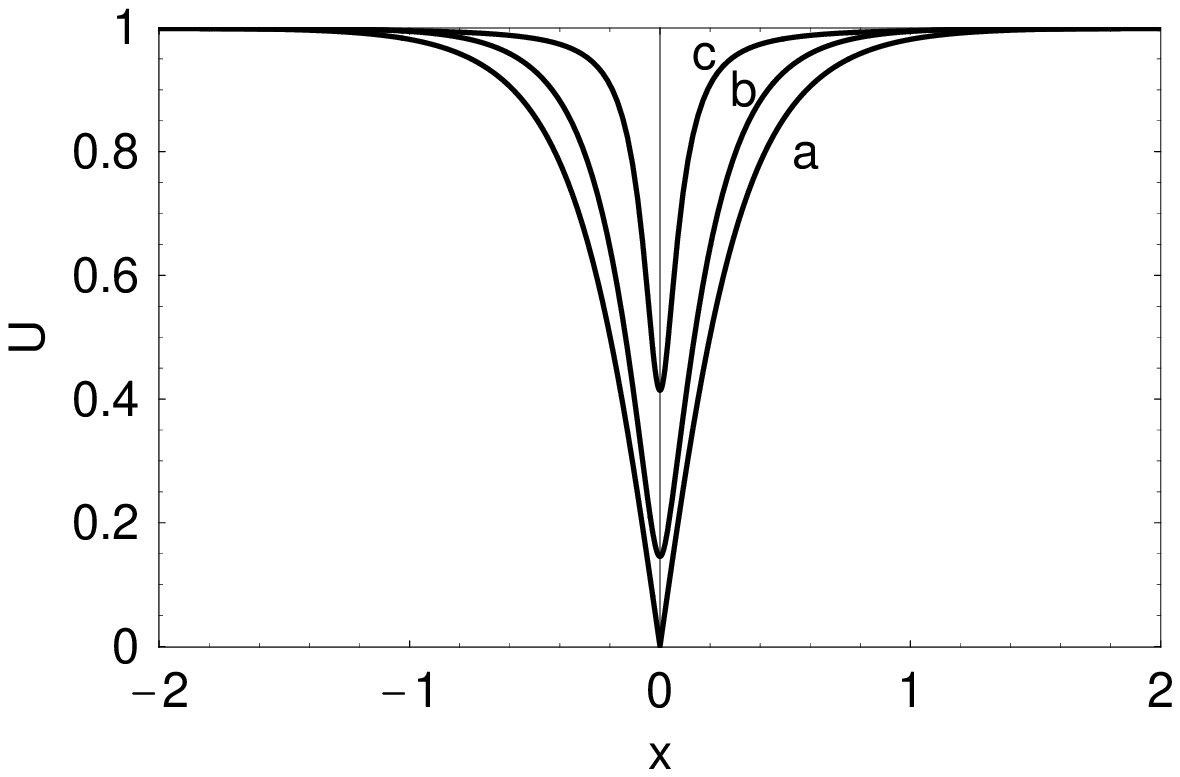}
\end{center}
{\bf Figure 7.} Profile of the amplitude of the dark soliton $U=|u_1|$ at $t=0$. a: $c=c_0=0.25$, b: $c=0.16$, c: $c=0.043$. The profile a is a black soliton and the profile c is an algebraic soliton.
\end{figure}
    \noindent {\it (i) Dark soliton: $a_1<0$}\par
 \medskip
 \noindent  It follows from (4.5) with $\kappa=-K, c_1=-c$ that $\kappa+b_1=1/2Kc-K/2$. Since $c_{\rm min}^\prime<c<c_{\rm max}^\prime$ by (4.23), the possible value of
 $\kappa+b_1$ is restricted by the inequality
 $$K\left[K\rho^2-1-\sqrt{K\rho^2(K\rho^2-1)}\right]<\kappa+b_1<K\left[K\rho^2-1+\sqrt{K\rho^2(K\rho^2-1)}\right]. \eqno(4.24)$$
   One can see that  the upper limit of $\kappa+b_1$ is attained when $c=c_{\rm min}^\prime$ and its limiting value is positive by the condition $K\rho^2>1$ 
   whereas the lower limit is attained when $c=c_{\rm max}^\prime$ and  is negative.
   In view of this fact, the algebraic dark soliton would be produced in the limit
 $c\rightarrow c_{\rm min}^\prime$ for which ${\rm sgn}(\kappa+b_1)>0$. Actually, taking the limit $a_1\rightarrow -0$ for the solutions (4.10) and (4.12) and using the expansion formulas (4.20), we find that
 the hyperbolic soliton reduces to the limiting form
 $$u_1=\rho\,{\rm e}^{{\rm i}(-Kx-\omega t)}\,{x-ct+x_0-{\rm i}\,{-2K+b_1\over 2b_1(-K+b_1)} \over x-ct+x_0-{\rm i}\,{1\over 2(-K+b_1)}}, \eqno(4.25a)$$
 $$|u_1|^2=\rho^2-{2K c^2\over 1-K^2c}\,{1\over (x-ct+x_0)^2+\left({K c\over 1-K^2c}\right)^2},\eqno(4.25b)$$
 where $b_1=(1+K^2c)/2K c$ and $c=c_{\rm min}^\prime$. 
 Since $1-Kc^\prime_{\rm min}>0$ by virtue of the condition  $K\rho^2>1$, the expression (4.25b) actually represents an algebraic dark soliton. 
 \par
 The black soliton appears when the velocity $c$ takes a specific value $c=c_0^\prime$, where 
 $$c_0^\prime=(K\rho^2-1)/K^2 \qquad {\rm at}\quad \gamma_1=\gamma_0^\prime=\cos^{-1}\left[{(K\rho^2)^{1\over 2}(3-2K\rho^2)\over 2(K\rho^2-1)^{3\over 2}}\right], \qquad (\pi<\gamma_0^\prime<2\pi). \eqno(4.26)$$
    Its profile is represented by  
  $$|u_1|^2=\rho^2\left[1-{3K\rho^2-4\over 2(K\rho^2-1)}{1\over \cosh\,2(\theta_1+\delta_1)+{K\rho^2-2\over 2(K\rho^2-1)}}\right]. \eqno(4.27)$$
  It is important to notice that the  inequality $c_{\rm min}^\prime<c_0^\prime<c_{\rm max}^\prime$ requires the condition $K\rho^2>4/3$ for the wavenumber $K$.
  It then turns out that  expression (4.27) takes the form of a black soliton. \par
    \begin{figure}[t]
\begin{center}
\includegraphics[width=10cm]{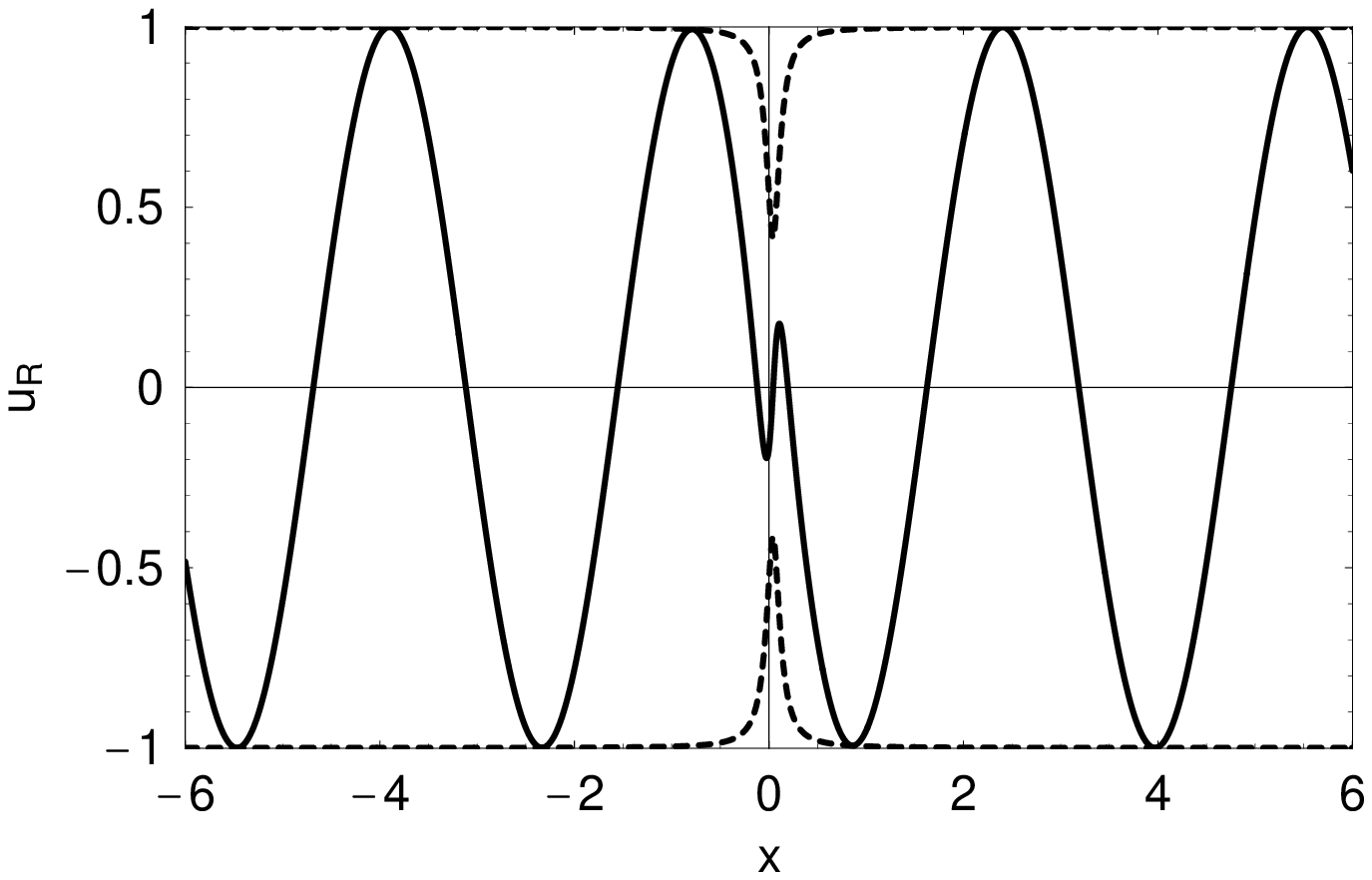}
\end{center}
\centerline{{\bf Figure 8.} Profile of an algebraic dark soliton $u_{\rm R}={\rm Re}\,u_1$ at $t=1$.} \par
\end{figure}
Figure 7 depicts
the profile of $U=|u_1|$ at $t=0$ for three different values of $c$, i.e., a: $c=c_0^\prime=0.25 (\gamma_1=\gamma_0^\prime=5\pi/4)$, b: $c=0.16 (\gamma_1=4\pi/3)$, c: $c=0.043 (\gamma_1=2\pi)$
with the parameters $\rho=1, \kappa=-2, \theta_{10}=-\delta_1$ and $\chi_{10}=0$. In this example, $c_{\rm min}^\prime=0.043, c_0^\prime=0.25$ and $c_{\rm max}^\prime=1.46$\ (see figure 6).
An algebraic soliton appears at  the lower limit of the velocity, i.e., $c=c_{\rm min}^\prime$ whereas a black soliton arises at $c=c_0^\prime$.
Figure 8 shows the profile of $u_R={\rm Re}\ u_1$ at $t=1$ for an algebraic dark soliton. \par
\medskip
 \noindent {\it (ii) Bright soliton: $a_1>0$}\par
 \medskip
 \begin{figure}[t]
\begin{center}
\includegraphics[width=10cm]{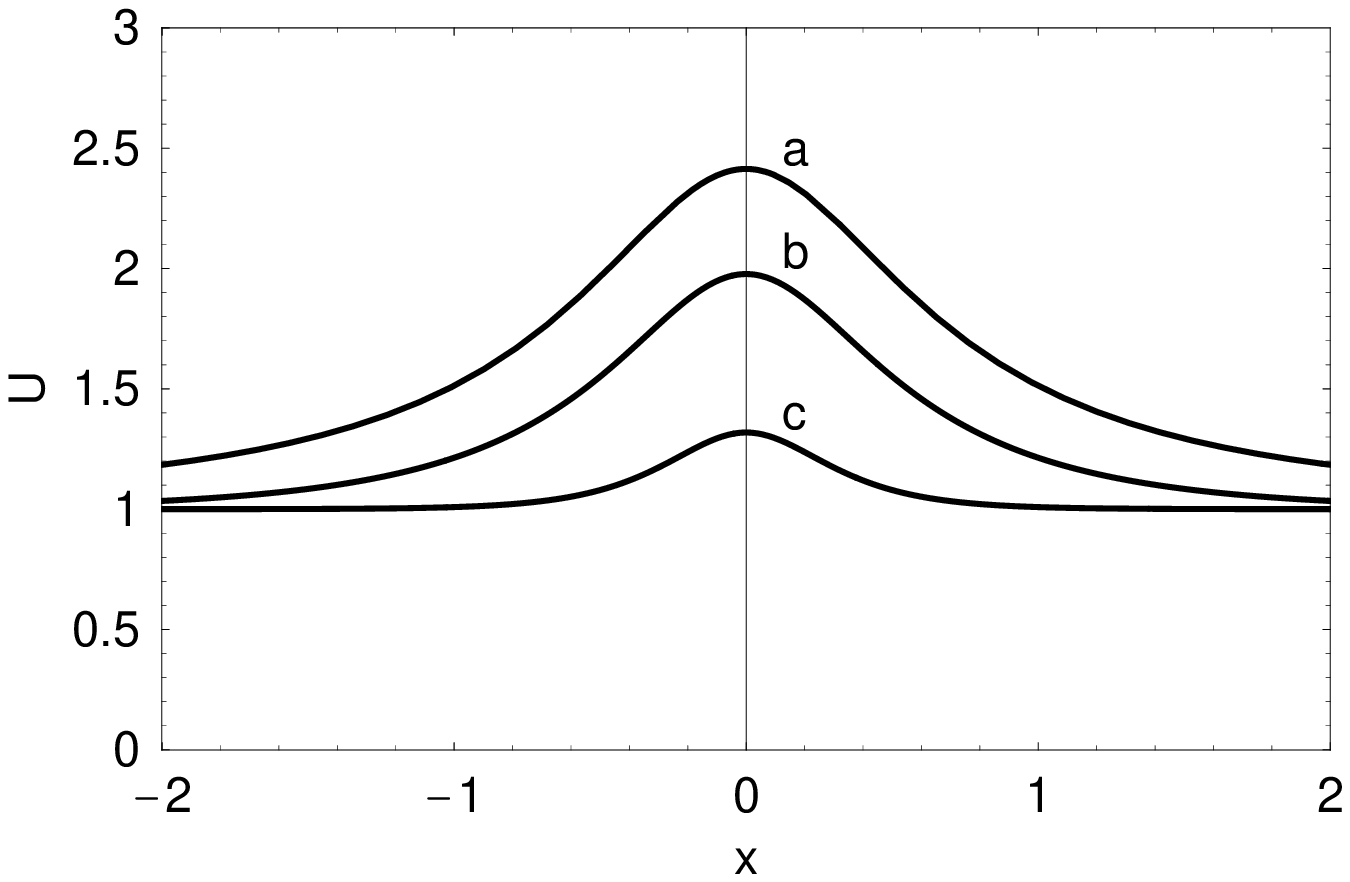}
\end{center}
{\bf Figure 9.} Profile of the amplitude of the bright soliton $U=|u_1|$ at $t=0$. a: $c=1.46$, b: $c=0.81$, c: $c=0.19$. The profiles  a is an algebraic soliton.\par
\end{figure}

\begin{figure}[t]
\begin{center}
\includegraphics[width=10cm]{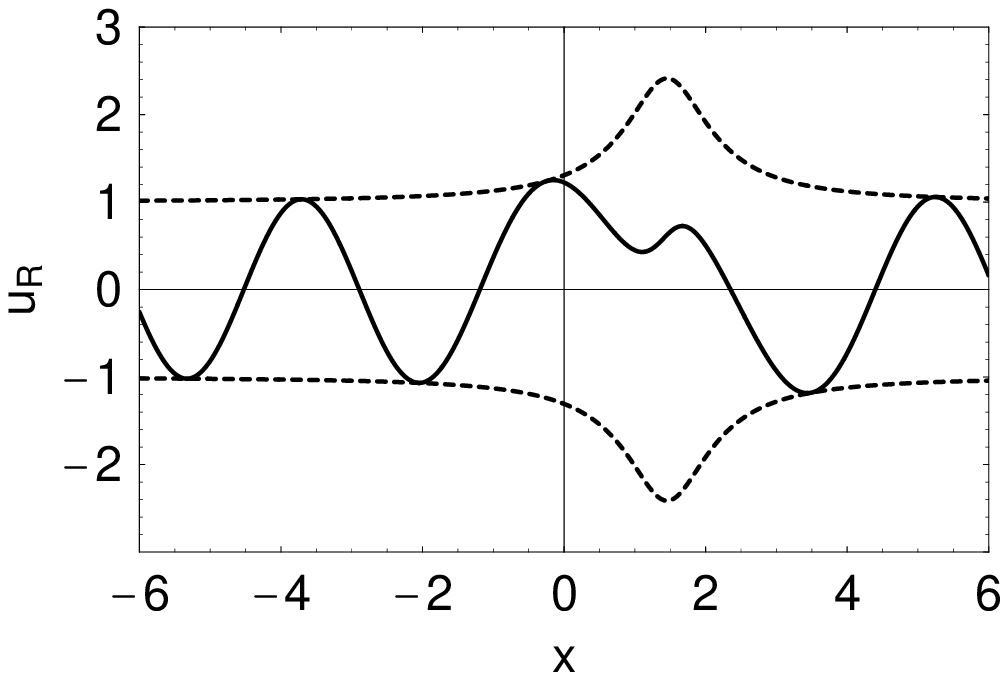}
\end{center}
{\bf Figure 10.} Profile of an algebraic bright soliton $u_R={\rm Re\, u_1}$  at $t=1$. \par
\end{figure}
 \noindent The crucial difference between the case 1 and the case 2 for the bright solitons is observed if one compares figure 6 with figure 1.
Notably, the bright soliton with $\kappa<0$ reduces to an algebraic soliton only at the upper limit of the velocity $c=c_{\rm max}^\prime$ whereas the bright soliton with $\kappa>0$
has two critical velocities $c_{\rm max}$ and  $c_{\rm min}$ for which algebraic solitons are produced. 
Figure 9 depicts the profile of $U=|u_1|$ at $t=0$ for three different values of $c$, i.e., a: $c=1.46 (\gamma_1=0.998\pi)$, b: $c=0.73 (\gamma_1=0.9\pi)$, c: $c= 0.025 (\gamma_1=0.7\pi)$
with $\rho=1$ and $\kappa=-2$.
Figure 10 shows the profile  $u_{\rm R}={\rm Re}\,u_1$ of an algebraic bright soliton at $t=1$ which corresponds to the profile a in figure 9. \par
\medskip
\noindent {\it 4.1.3.  Note on algebraic solitons }\par
\medskip
\noindent We have seen that the algebraic solitons arise from the hyperbolic solitons when certain conditions are satisfied. 
Here, we summarize the result. The algebraic bright solitons are produced when the conditions ${\rm sgn}\,a_1{\rm sgn}(\kappa+b_1)=-1$ and ${\rm sgn}(\kappa a_1)<0$ are
satisfied simultaneously whereas the corresponding conditions for the dark algebraic solitons are given by ${\rm sgn}\,a_1{\rm sgn}(\kappa+b_1)=-1$ and ${\rm sgn}(\kappa a_1)>0$.
Thus, for $\kappa>0$, the conditions ${\rm sgn}(\kappa+b_1)=1$ and  ${\rm sgn}(\kappa+b_1)=-1$ are responsible for the generation of the algebraic
bright and dark solitons, respectively. Since $\kappa+b_1>0$ in this case, only the bright algebraic soliton exists.  See figure 1.
For $\kappa<0$, on the other hand, the above
conditions turn out to be ${\rm sgn}(\kappa+b_1)=-1$ and  ${\rm sgn}(\kappa+b_1)=1$, respectively. Under this setting, the limiting value of $\kappa+b_1$ becomes negative
for the bright soliton and positive for the dark soliton, respectively, implying the existence of both types of algebraic solitons.  See figure 6.
In conclusion, we emphasize that
the criterion for the existence of solitons (which depends crucially on the sign of $\kappa$) plays an important role in our analysis. \par
\bigskip
\noindent{\it 4.2. Two-soliton solution}\par
\medskip
 \noindent As clarified by the analysis of the ono-soliton solutions, both dark and bright solitons exist in our system. 
 Therefore, the two-soliton solutions can be classified into three types, i.e., dark-dark solitons, dark-bright solitons and bright-bright solitons.
 Here, we  
focus our attention on the dark-dark  solitons. Especially, we investigate the asymptotic behavior of the solution for large time.  
The two-soliton solution describing the interaction between a dark soliton and a bright soliton will be briefly discussed. 
For both cases, we assume that $\kappa>0$.  
\par
\medskip
\noindent{\it 4.2.1.  Dark-dark solitons}\par
\medskip
\noindent The tau functions
$f_2$ and $g_2$ representing the dark two-soliton solution are given by (3.1)-(3.3) with $N=2$ subjected to the conditions $\kappa>0, a_1>0, a_2>0$. They read
$$f_2=1+{\kappa-{\rm i}p_1\over p_1+p_1^*}\,z_1z_1^*+{\kappa-{\rm i}p_2\over p_2+p_2^*}\,z_2z_2^*
+{(\kappa-{\rm i}p_1)(\kappa-{\rm i}p_2)(p_1-p_2)(p_1^*-p_2^*)\over (p_1+p_1^*)(p_1+p_2^*)(p_2+p_1^*)(p_2+p_2^*)}\,z_1z_2z_1^*z_2^*, \eqno(4.28a)$$
$$g_2\!=\!1\!-\!{\kappa+{\rm i}p_1^*\over p_1+p_1^*}{p_1\over p_1^*}\,z_1z_1^*\!-\!{\kappa+{\rm i}p_2^*\over p_2+p_2^*}{p_2\over p_2^*}\,z_2z_2^*
\!+\!{(\kappa+{\rm i}p_1^*)(\kappa+{\rm i}p_2^*)(p_1-p_2)(p_1^*-p_2^*)\over (p_1+p_1^*)(p_1+p_2^*)(p_2+p_1^*)(p_2+p_2^*)}{p_1p_2\over p_1^*p_2^*}\,z_1z_2z_1^*z_2^*. \eqno(4.28b)$$
\par
To investigate the interaction process of two solitons, 
 we first order the magnitude of the velocity of each soliton in the $(x, t)$ coordinate system as $c_1>c_2>0$.  Invoking the definition (4.2a) of the velocity of the solitons, this can be established by
imposing the condition $|p_1|<|p_2|$ on the amplitude parameters. 
Now, we take the limit $t\rightarrow -\infty$  with $\theta_1$ being fixed. 
Since in this limit $|z_1|=$finite and $|z_2|\rightarrow \infty$, the leading-order asymptotics of $f_2$ and $g_2$ are found to be as
$$f_2 \sim {\kappa-{\rm i}p_2\over p_2+p_2^*}\,z_2z_2^*\left\{1+{(\kappa-{\rm i}p_1)(p_1-p_2)(p_1^*-p_2^*)\over (p_1+p_1^*)(p_1+p_2^*)(p_2+p_1^*))}\,z_1z_1^*\right\}, \eqno(4.29a)$$
$$g_2 \sim -{\kappa+{\rm i}p_2^*\over p_2+p_2^*}{p_2\over p_2^*}\,z_2z_2^*\left\{1-{(\kappa+{\rm i}p_1^*)(p_1-p_2)(p_1^*-p_2^*)\over (p_1+p_1^*)(p_1+p_2^*)(p_2+p_1^*)}{p_1\over p_1^*}\,z_1z_1^*\right\}. \eqno(4.29b)$$
The asymptotic form of the two-dark soliton solution follows from (2.1) upon substituting (4.29) into it, giving rise to
$$u_2 \sim \rho\,{\rm exp}\left\{{\rm i}\left(\kappa x-\omega t+\phi_1^{(-)}\right)\right\}{1-{\kappa+{\rm i}p_1^*\over p_1+p_1^*}{p_1\over p_1^*}\,z_1^\prime {z_1^\prime}^* \over 1
+{\kappa-{\rm i}p_1\over p_1+p_1^*}\,z_1^\prime {z_1^\prime}^*}, \eqno(4.30a)$$
where
$$z_1^\prime=z_1\,{\rm exp}\left[-{\rm ln}\left({p_1+p_2^*\over p_1-p_2}\right)\right], \eqno(4.30b)$$
$$\phi_1^{(-)}={\rm arg}\left({\kappa+{\rm i}p_2^*\over \kappa-{\rm i}p_2}{p_2\over p_2^*}\right)+\pi. \eqno(4.30c)$$
Let $u_1(\theta_1)$ be the dark one-soliton solution (4.10). Then, the asymptotic form of $u_2$ can be written in terms of $u_1$ as
$$u_2 \sim {\rm exp}\left({\rm i}\phi_1^{(-)}\right)u_1(\theta_1+\Delta \theta_1^{(-)}), \quad  \Delta\theta_1^{(-)}=-{\rm ln}\left|{p_1+p_2^*\over p_1-p_2}\right|. \eqno(4.31)$$
\par
Next, we take the limit $t\rightarrow +\infty$ with $\theta_1$ being fixed. In this limit, $|z_1|=$finite and $|z_2|\rightarrow 0$. Therefore, the tau functions $f_2$ and $g_2$ and
the two-soliton solution $u_2$
behave like
$$f_2\sim 1+{\kappa-{\rm i}p_1\over p_1+p_1^*}\,z_1z_1^*,\qquad g_2\sim 1-{\kappa+{\rm i}p_1^*\over p_1+p_1^*}{p_1\over p_1^*}\,z_1z_1^*, \eqno(4.32)$$
$$u_2 \sim \rho\,{\rm e}^{{\rm i}(\kappa x-\omega t)}\,{1-{\kappa+{\rm i}p_1\over p_1+p_1^*}{p_1\over p_1^*}\,z_1z_1^* \over 1+{\kappa-{\rm i}p_1\over p_1+p_1^*}\,z_1z_1^*}. \eqno(4.33)$$
It follows from (4.33) that 
$$u_2 \sim u_1(\theta_1+\Delta \theta_1^{(+)}), \quad  \Delta\theta_1^{(+)}=0. \eqno(4.34)$$
\par
The trajectory of the center position $x=x_c(t)$ of the $j$th soliton is described by the equation $\theta_j+\Delta\theta_j^{(\pm)}=0$, or $x_c=-c_jt-(\theta_{j0}+\Delta\theta_j^{(\pm)})/a_j$.
Since the soliton propagates to the left, the phase shift $\Delta x_j$ of the $j$th soliton can be defined by the relation
$$\Delta x_j=x_c(-\infty)-x_c(+\infty)={1\over a_j}\left(\Delta\theta_j^{(+)}-\Delta\theta_j^{(-)}\right), \quad j=1, 2. \eqno(4.35)$$
We see from (4.31) and (4.34) that the fast soliton suffers a phase shift
$$\Delta x_1={1\over a_1}{\rm ln}\left|{p_1+p_2^*\over p_1-p_2}\right|. \eqno(4.36)$$
In terms of the angular variable $\gamma_1$ and $\gamma_2$ defined by (4.6) and (4.7), this expression can be rewritten in the form
$$\Delta x_1={1\over a_1}\,{\rm ln}\left|{\sin{1\over 2}\left(\gamma_1+\gamma_2\right)\over \sin{1\over 2}\left(\gamma_1-\gamma_2\right)}\right|,
\quad a_1=\sqrt{\kappa^3\rho^2(1+\kappa\rho^2)}\,\sin\,\gamma_1,\qquad 0<\gamma_1<\pi. \eqno(4.37)$$
\par
We can perform the similar asymptotic analysis while keeping $\theta_2$ fixed. Hence, we quote only the final results. As $t\rightarrow -\infty$, the
expressions corresponding to (4.29) and (4.31) read respectively
$$f_2\sim 1+{\kappa-{\rm i}p_2\over p_2+p_2^*}\,z_2z_2^*,\qquad g_2\sim 1-{\kappa+{\rm i}p_2^*\over p_2+p_2^*}{p_2\over p_2^*}\,z_2z_2^*, \eqno(4.38)$$
$$u_2 \sim u_1(\theta_2+\Delta \theta_2^{(-)}), \quad  \Delta\theta_2^{(-)}=0. \eqno(4.39)$$
As $t\rightarrow \infty$, on the other hand, they take the form
$$f_2 \sim {\kappa-{\rm i}p_1\over p_1+p_1^*}\,z_1z_1^*\left\{1+{(\kappa-{\rm i}p_2)(p_1-p_2)(p_1^*-p_2^*)\over (p_2+p_2^*)(p_1+p_2^*)(p_2+p_1^*))}\,z_2z_2^*\right\}, \eqno(4.40a)$$
$$g_2 \sim -{\kappa+{\rm i}p_1\over p_1+p_1^*}{p_1\over p_1^*}\,z_1z_1^*\left\{1-{(\kappa+{\rm i}p_2^*)(p_1-p_2)(p_1^*-p_2^*)\over (p_2+p_2^*)(p_1+p_2^*)(p_2+p_1^*))}{p_2\over p_2^*}\,z_2z_2^*\right\}, \eqno(4.40b)$$
$$u_2 \sim {\rm exp}\left({\rm i}\phi_2^{(+)}\right)u_1(\theta_2+\Delta \theta_2^{(+)}),\eqno(4.41a) $$
$$ \Delta\theta_2^{(+)}=-{\rm ln}\left|{p_2+p_1^*\over p_2-p_1}\right|, \quad
\phi_2^{(+)}={\rm arg}\left({\kappa+{\rm i}p_1^*\over \kappa-{\rm i}p_1}{p_1\over p_1^*}\right)+\pi. \eqno(4.41b)$$
\begin{figure}[t]
\begin{center}
\includegraphics[width=10cm]{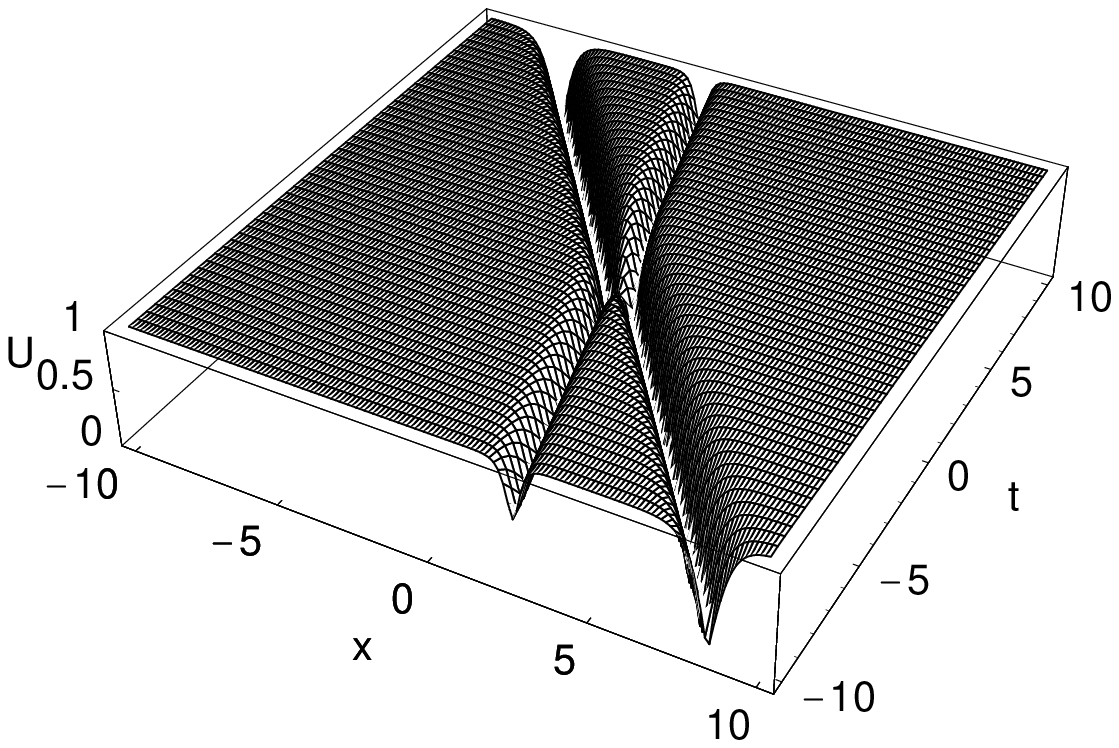}
\end{center}
\centerline{{\bf Figure 11.} The interaction of two dark solitons.}\par
\end{figure}
The phase shift of the slow soliton follows from (4.35), (4.39) and (4.41), resulting in
$$\Delta x_2=-{1\over a_2}{\rm ln}\left|{p_2+p_1^*\over p_2-p_1}\right|, \eqno(4.42)$$
or equivalently in terms of the angular variables $\gamma_1$ and $\gamma_2$, it reads
$$\Delta x_2=-{1\over a_2}\,{\rm ln}\left|{\sin{1\over 2}\left(\gamma_2+\gamma_1\right)\over \sin{1\over 2}\left(\gamma_2-\gamma_1\right)}\right|,
\quad a_2=\sqrt{\kappa^3\rho^2(1+\kappa\rho^2)}\sin\,\gamma_2, \qquad 0<\gamma_2<\pi. \eqno(4.43)$$
An inspection of the formulas (4.36) and (4.42) reveals that $\Delta x_1>0$ and $\Delta x_2<0$ under the setting $a_1>0,\ a_2>0$.
\par
Figure 11 shows the intercaction of two dark solitons 
with the parameters  $\rho=1, \kappa=2, c_1=0.75(\gamma_1=0.90\pi), c_2=0.24(\gamma_2=0.80\pi)$ and $\zeta_{10}=\zeta_{20}=0$ so that from (4.14), $A_{d1}=1.0$ and $A_{d2}=0.47$.
It can be seen from figure 1 that the amplitude of each dark soliton is an increasing function of the velocity for the present choice of the parameters.
Note, in this example, that the large soliton is a black soliton since its asymptotic amplitude is $A_{d1}=\rho=1$. 
The phase shifts evaluated from the formulas (4.37) and (4.43) are given by $\Delta x_1=0.70$ and $\Delta x_2=-0.36$, respectively. 
 Figure 11 shows clearly a typical interaction process of solitons, i.e., as time goes, the large soliton gets close to the small soliton and overtakes it 
and after the collision,
both solitons eventually separate each other  without changing their profiles. The net effect of the collision is only  the phase shift. \par
\medskip
\noindent{\it 4.2.2.  Dark-bright solitons}\par
\medskip
\noindent The two-soliton solution consisting of a dark soliton and a bright soliton is obtained by choosing the parameters such as $\kappa>0, a_1>0$ and $a_2<0$, for example. 
The asymptotic analysis can be performed as well for this solution and hence the detail will be omitted. 
\begin{figure}[t]
\begin{center}
\includegraphics[width=10cm]{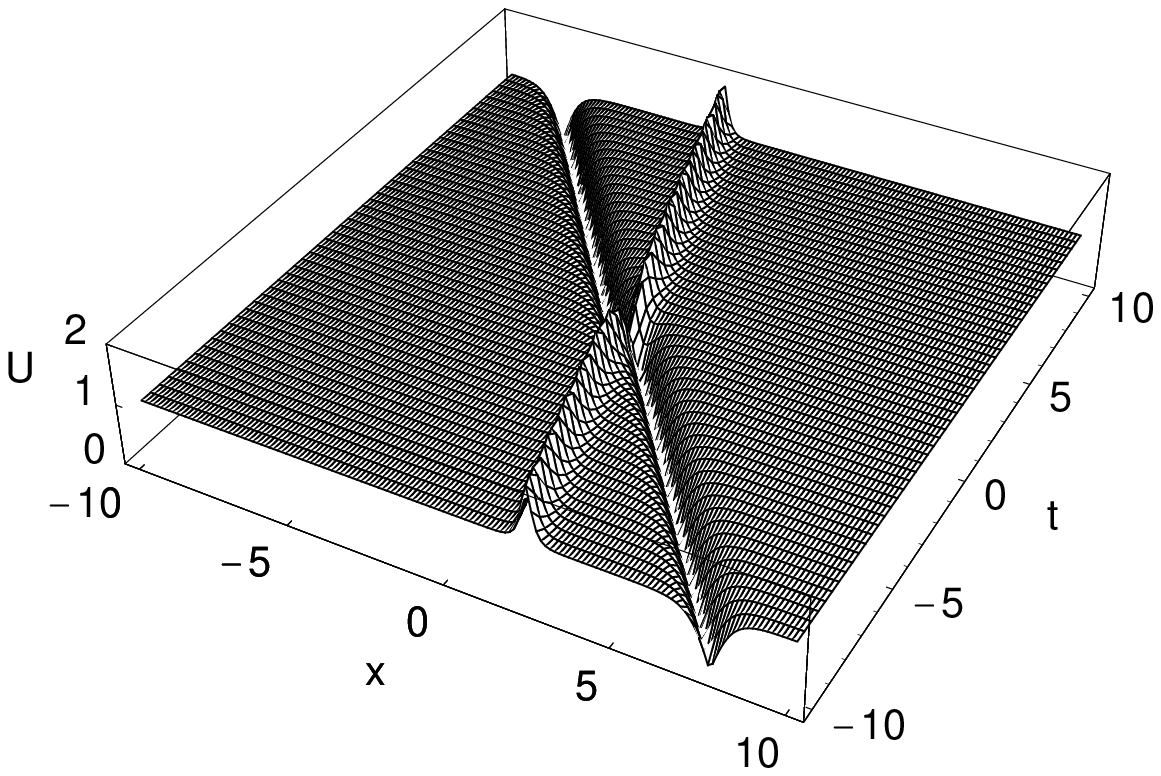}
\end{center}
\centerline{{\bf Figure 12.} The interaction between a dark soliton and a bright soliton.}\par
\end{figure}
\par
Figure 12 depicts the interaction between a dark soliton and a bright soliton
with the parameters $\rho=1, \kappa=2, c_1=0.75(\gamma_1=0.90\pi), c_2=0.24(\gamma_2=1.2\pi)$ and $\zeta_{10}=\zeta_{20}=0$, showing that
the dark soliton propagates faster than the bright soliton.
The asymptotic amplitudes of the dark and bright solitons are given respectively by
$A_{d1}=1.0$ and $A_{b2}=0.92$ and hence the former is a black soliton.
The figure clearly shows the solitonic behavior of the solution. The dark soliton suffers a positive phase shift whereas the bright soliton suffers a negative phase shift.
The formulas $\Delta x_1$ for the dark soliton and $\Delta x_2$ for the bright soliton for the phase shifts are given respectively by 
$$\Delta x_1=-{1\over a_1}\,{\rm ln}\left|{\sin{1\over 2}\left(\gamma_1+\gamma_2\right)\over \sin{1\over 2}\left(\gamma_1-\gamma_2\right)}\right|,
\quad a_1=\sqrt{\kappa^3\rho^2(1+\kappa\rho^2)}\,\sin\,\gamma_1,\qquad 0<\gamma_1<\pi,  \eqno(4.44a)$$
$$\Delta x_2=-{1\over a_2}\,{\rm ln}\left|{\sin{1\over 2}\left(\gamma_2+\gamma_1\right)\over \sin{1\over 2}\left(\gamma_2-\gamma_1\right)}\right|,
\quad a_2=\sqrt{\kappa^3\rho^2(1+\kappa\rho^2)}\sin\,\gamma_2, \qquad \pi<\gamma_2<2\pi. \eqno(4.44b)$$
As in the case of the dark-dark solitons, one can see that $\Delta x_1>0$ and $\Delta x_2<0$.
In the present example, $\Delta x_1= 0.70$ and $\Delta x_2=-0.36$. \par
\bigskip
\noindent{\it 4.3. Dark $N$-soliton solution}\par
\medskip
 \noindent The preceding  analysis reveals that
 the asymptotic form of the $N$-soliton solution will be represented by a superposition of  $n$ dark solitons and $N-n$ bright solitons 
where $n$ is an arbitrary nonnegative integer in the interval $0\leq n\leq N$. 
 The derivation of the large time asymptotic  for the general $N$-soliton solution can be done following the similar procedure to that used for the
 two-soliton case. Hence, we outline the result. We address the dark soliton solutions satisfying the conditions $\kappa>$ and $a_j>0\ (j=1, 2, ..., N)$.
 The analysis for the bright soliton solutions as well as 
 an arbitrary combination of dark and bright solitons can be carried out in exactly the  same way. \par
To begin with, we order the magnitude of the velocity of each soliton  as $c_1>c_2> ... >c_N>0$. We take the limit $t \rightarrow -\infty$ with $\theta_n$ being finite. 
Since in this limit, $|z_j|\rightarrow 0$ for $j<n$ and $|z_j|\rightarrow \infty$ for $n<j$, we find that the leading-order asymptotic of
the tau function $f=f_N$  from (3.1) with (3.2)  can be written in the form
$$f_N\sim \left|(c_{jk})_{n+1\leq j,k\leq N}\right|\prod_{j=n+1}^N(z_jz_j^*)\left(1+{\left|(c_{jk})_{n\leq j,k\leq N}\right|\over \left|(c_{jk})_{n+1\leq j,k\leq N}\right|}\,z_nz_n^*\right). \eqno (4.45a)$$
Here, $(c_{jk})$ is a matrix of Cauchy type given by
$$c_{jk}={\kappa-{\rm i}p_j\over p_j+p_k^*}, \quad 1\leq j,\ k\leq N. \eqno(4.45b)$$
Referring to the well-known Cauchy's formula, the determinant of the  matrix $(c_{jk})$ is evaluated as
$$\left|(c_{jk})_{m\leq j,k\leq n}\right|=\prod_{j=m}^n(\kappa-{\rm i}p_j){\prod_{m\leq j<k\leq n}(p_j-p_k)(p_j^*-p_k^*)\over \prod_{m\leq j,k\leq n}(p_j+p_k^*)},\quad 1\leq m<n\leq N. \eqno(4.45c)$$
If we use (4.45c), we have
$${\left|(c_{jk})_{n\leq j,k\leq N}\right|\over \left|(c_{jk})_{n+1\leq j,k\leq N}\right|}={\kappa-{\rm i}p_n\over p_n+p_n^*}\,{\rm exp}\left[-\sum_{j=n+1}^N{\rm ln}\left({p_n+p_j^*\over p_n-p_j}\right)
-\sum_{j=n+1}^N{\rm ln}\left({p_n^*+p_j\over p_n^*-p_j^*}\right)\right]. \eqno(4.46)$$
Substitution of (4.46) into (4.45) now gives 
$$f_N\sim \left|(c_{jk})_{n+1\leq j,k\leq N}\right|\prod_{j=n+1}^N(z_jz_j^*)\left(1+{\kappa-{\rm i}p_n\over p_n+p_n^*}\,z_n^\prime {z_n^\prime}^*\right), \eqno(4.47a)$$
where
$$z_n^\prime=z_n\,{\rm exp}\left[-\sum_{j=n+1}^N{\rm ln}\left({p_n+p_j^*\over p_n-p_j}\right)\right]. \eqno(4.47b)$$
The leading-order asymptotic of $g_N$ in the limit of $t\rightarrow -\infty$ can be derived in the same way. It takes the form
$$g_N\sim \left|(c_{jk}^\prime)_{n+1\leq j,k\leq N}\right|\prod_{j=n+1}^N(z_jz_j^*)\left(1-{\kappa+{\rm i}p_n^*\over p_n+p_n^*}\,{p_n\over p_n^*}\,z_n^\prime {z_n^\prime}^*\right), \eqno(4.48a)$$
where
$$c_{jk}^\prime=-{\kappa-{\rm i}p_j\over p_j+p_k^*}\,{p_j\over p_k^*}, \quad 1\leq j, k\leq N. \eqno(4.48b)$$
The asymptotic form of the dark $N$-soliton solution follows from (2.1), (4.47) and (4.48). It reads
$$u_N\sim \rho\,{\rm exp}\left\{{\rm i}
\left(\kappa x-\omega t+\phi_n^{(-)}\right)\right\}{1-{\kappa+{\rm i}p_n^*\over p_n+p_n^*}\,{p_n\over p_n^*}\,z_n^\prime {z_n^\prime}^*\over 1+{\kappa-{\rm i}p_n\over p_n+p_n^*}\,z_n^\prime {z_n^\prime}^*}, \eqno(4.49a)$$
with
$$\phi_n^{(-)}={\rm arg}\left[\prod_{j=n+1}^N\left({\kappa+{\rm i}p_j^*\over \kappa-{\rm i}p_j}\,{p_j\over p_j^*}\right)\right]+(N-n)\pi. \eqno(4.49b)$$
This expression can be rewritten in terms of the one-soliton solution as
$$u_N\sim {\rm exp}\left({\rm i}\phi_n^{(-)}\right)u_1(\theta_n+\Delta\theta_n^{(-)}), \eqno(4.50a)$$
with 
$$\Delta\theta_n^{(-)}=-\sum_{j=n+1}^N{\rm ln}\left|{p_n+p_j^*\over p_n-p_j}\right|. \eqno(4.50b)$$
\par
By a similar asymptotic analysis, we can derive the asymptotic form of $u_N$ in the limit of $t\rightarrow +\infty$. We find that
$$u_N\sim {\rm exp}\left({\rm i}\phi_n^{(+)}\right)u_1(\theta_n+\Delta\theta_n^{(+)}), \eqno(4.51a)$$
with 
$$\Delta\theta_n^{(+)}=-\sum_{j=1}^{n-1}{\rm ln}\left|{p_n+p_j^*\over p_n-p_j}\right|, \eqno(4.51b)$$
$$\phi_n^{(+)}={\rm arg}\left[\prod_{j=1}^{n-1}\left({\kappa+{\rm i}p_j^*\over \kappa-{\rm i}p_j}\,{p_j\over p_j^*}\right)\right]+(n-1)\pi. \eqno(4.51c)$$
We see from (4.50) and (4.51) that in the rest frame of reference, the asymptotic form of the dark $N$-soliton solution can be represented by a superposition of $N$ independent
dark one-soliton solutions, the only difference being the phase shifts of each soliton caused by the collisions.
It follows from (4.50b) and (4.51b) that the formula for the total phase shift of the $n$th soliton is given by 
$$\Delta x_n={1\over a_n}\left(\sum_{j=n+1}^N{\rm ln}\left|{p_n+p_j^*\over p_n-p_j}\right|-\sum_{j=1}^{n-1}{\rm ln}\left|{p_n+p_j^*\over p_n-p_j}\right|\right), \quad n=1, 2, ..., N. \eqno(4.52)$$
As in the two-soliton case, we can rewrite the above formula in terms of the variables $\gamma_j$ defined by (4.6) and (4.7). Explicitly,
$$\Delta x_n={1\over a_n}\left(\sum_{j=n+1}^N{\rm ln}\left|{\sin{1\over 2}(\gamma_n+\gamma_j)\over \sin{1\over 2}(\gamma_n-\gamma_j)}\right|
-\sum_{j=1}^{n-1}{\rm ln}\left|{\sin{1\over 2}(\gamma_n+\gamma_j)\over \sin{1\over 2}(\gamma_n-\gamma_j)}\right|\right), $$
$$a_n=\sqrt{\kappa^3\rho^2(1+\kappa\rho^2)}\sin\,\gamma_n, \qquad 0<\gamma_n<\pi,\qquad n=1, 2, ..., N. \eqno(4.53)$$
The  formulas (4.52) and (4.53) reduce to (4.36), (4.37), (4.42) and (4.43) for the special case of $N=2$. They clearly show that each soliton has pairwise interactions with
other solitons, i.e., there are no many-particle collisions among solitons. This feature is common to that of the bright $N$-soliton solution considered in I. \par
\bigskip
\leftline{\bf  5. Concluding remarks} \par
\bigskip
\noindent In this paper, the system of bilinear equations reduced from the  FL equation has been derived and used
to construct the dark $N$-soliton solution. 
The corresponding $N$-soliton solution derived in [7] using the B\"acklund transformation follows from our solution (2.1) with (3.1) and (3.2) if one
introduces the angular variables $\gamma_j$ according to the relations (4.7).  We have found that unlike the bright soliton solutions obtained in I, the
complex amplitude parameters $p_j$ are subjected to the constraints (3.2c) which have prevented
the proof of  the solution. To overcome this difficulty, we have employed a trilinear equation in place of one of the bilinear equations, in addition to an auxiliary variable $\tau$ in (3.2c).
As a byproduct,  this trilinear equation has led for the first time to a simple formula for the dark $N$-soliton solution of the derivative NLS equation on the background of a plane wave.
Note that  the dark soliton solutions on a constant background [13-19] stem simply from the above-mentioned solution in the zero limit of the wavenumber $\kappa$. However, this limiting procedure
is found to be  unable to perform for the dark $N$-soliton solution of the FL equation due to the singularity of the dispersion relation. \par
We have seen that the soliton solutions presented here exhibit several new features. Specifically, both the dark and bright solitons
exist depending on  the sign of the wavenumber $\kappa$ and that of the real part of the complex amplitude parameter.
Of particular interest is the existence of an algebraic dark soliton which appears only in the case of negative $\kappa$.
Finally, the asymptotic analysis of the two- and general $N$-soliton solutions has clarified their structure and dynamics. In particular, the latter solution has been shown to
include $n$ dark solitons and $N-n$ bright solitons on nonzero background with $n$ being an arbitrary nonnegative integer not exceeding $N$.
The application of the results summarized above to nonlinear fiber optics 
 will be an interesting issue to be studied in a future research work. 
\par
\bigskip
\leftline{\bf Acknowledgement}\par
\bigskip
\noindent This work was partially supported by the Grant-in-Aid for Scientific Research (C) No. 22540228 from Japan Society for the Promotion of Science. \par

\newpage

\leftline{\bf References}\par
\begin{enumerate}[{[1]}]
\item Fokas A S 1995 {On a class of physically important integrable equations} {\it Physica D} {\bf 87} 145-150
\item Lenells J 2009 {Exactly solvable model for nonlinear pulse propagation in optical fibers} {\it Stud. Appl. Math.} {\bf 123} 215-232
\item Lenells J and Fokas A S 2009 {On a novel integrable generalization of the nonlinear Schr\"odinger equation} {\it Nonlinearity} {\bf 22} 11-27
\item Lenells J 2010 {Dressing for a novel integrable generalization of the nonlinear Schr\"odinger equation} {\it J. Nonlinear Sci.} {\bf 20} 709-722
\item Kundu A 2010 {Two-fold integrable hierarchy of nonholonomic deformation of the derivative nonlinear Schr\"odinger and the Lenells-Fokas equation} {\it J. Math. Phys.} {\bf 51} 022901
\item Matsuno Y 2011 {A direct method of solution for the Fokas-Lenells derivative nonlinear Schr\"odinger equation: I. Bright soliton solutions} {\it J. Phys. A: Math. Theor.} {\bf 45} 235202
\item Vekslerchik V E 2011 {Lattice representation and dark solitons of the Fokas-Lenells equation} {\it Nonlinearity} {\bf 24} 1165-1175
\item Hirota R 2004 {\it The Direct Method in Soliton Theory} (New York: Cambridge)
\item Matsuno Y 1984 {\it Bilinear Transformation Method} (New York: Academic)
\item Vein R  and Dale P 1999 {\it Determinants and Their Applications in Mathematical Physics} (New York: Springer)
\item Matsuno Y 2011 {The {\it N}-soliton solution of a two-component modified nonlinear Schr\"odinger equation} {\it Phys. Lett.} {\bf A 375} 3090-3094
\item Matsuno Y 2011 {The bright {\it N}-soliton solution of a multi-component modified nonlinear Schr\"odinger equation} {\it J. Phys. A: Mat. Theor.} {\bf 44} 495202
\item Kawata T and Inoue H 1978 {Exact solutions of derivative nonlinear Schr\"odinger equation under the nonvanishing conditions} {\it J. Phys. Soc. Jpn.} {\bf 44} 1968-1976
\item Kawata T, Kobayashi N and Inoue H 1979 {Soliton solutions of the derivative nonlinear Schr\"odinger equation} {\it J. Phys. Soc. Jpn.} {\bf 46} 1008-1015
\item Chen X J, Yang J and Lam W K 2006 {$N$-soliton solution for the derivative nonlinear Schr\"odinger equation with nonvanishing boundary conditions} {\it J. Phys. A: Math. Gen.} {\bf 39} 3263-3274
\item Laskin V M 2007 {$N$-soliton solutions and perturbation theory for the derivative nonlinear Schr\"odinger equation with nonvanishing boundary condition} {\it J. Phys. A: Math. Theor.} {\bf 40} 6119-6132
\item Steudel H 2003 { The hierarchy of multi-soliton solutions of the derivative nonlinear Schr\"odinger equation} {\it J. Phys. A: Math. Gen.} {\bf 36} 1931-1946
\item Xu S, He J and Wang L 2011 {The Darboux transformation of the derivative nonlinear Schr\"odinger equation} {\it J. Phys. A: Math. Theor.} {\bf 44} 305203
\item Li M, Tian B, Liu W J, Zhang H Q and Wang P 2010 {Dark and antidark solitons in the modified nonlinear Schr\"odinger equation accounting for the self-steepening effect} {\it Phys. Rev. E} {\bf 81} 046606
\item Mij{\o}lhus E 1976 {On the modulational instability of hydromagnetic waves parallel to the magnetic field} {\it J. Plasma Phys.} {\bf 16} 321-334
\item Ichikawa Y H, Konno K, Wadati M and Sanuki H 1980 {Spiky soliton in circular polarized Alfv\'en wave} {\it J. Phys. Soc. Jpn.} {\bf 48} 279-286
\item Wright III O C 2009 {Some homoclinic connections of a novel integrable generalized nonlinear Schr\"odinger equation} {\it Nonlinearity} {\bf 22} 2633-2643
\item L\"u X and Tian B 2012 {Novel behavior and properties for the nonlinear pulse propagation in optical fibers} {\it Europhys. Lett.} {\bf 97} 10005
\item Mio K, Ogino T, Minami K and Takeda S 1976 {Modified nonlinear Schr\"odinger equation for Alfv\'en waves propagating along the magnetic field in cold plasmas} {\it J. Phys. Soc. Jpn.} {\bf 41} 265-271
\item Mij{\o}lhus E 1978 {A note on the modulational instability of long Alfv\'en waves parallel to the magnetic field} {\it J. Plasma Phys.} {\bf 19} 437-447

\end{enumerate}

\end{document}